# Title: Primate specific retrotransposons, SVAs, in the evolution of networks that alter brain function.

**Olga Vasieva[1*], Sultan Cetiner[1], Abigail Savage[2], Gerald G. Schumann[3], Vivien J Bubb[2], John P Quinn[2*],**

[1] Institute of Integrative Biology, University of Liverpool, Liverpool, L69 7ZB, U.K
[2] Department of Molecular and Clinical Pharmacology, Institute of Translational Medicine, The University of Liverpool, Liverpool L69 3BX, UK

[3] Division of Medical Biotechnology, Paul-Ehrlich-Institut, Langen, D-63225 Germany

*. Corresponding author

Olga Vasieva: Institute of Integrative Biology, Department of Comparative genomics, University of Liverpool, Liverpool, L69 7ZB, ovasieva@liverpool.ac.uk ; Tel: (+44) 151 795 4456; FAX:(+44) 151 795 4406

John Quinn: Department of Molecular and Clinical Pharmacology, Institute of Translational Medicine, The University of Liverpool, Liverpool L69 3BX, UK, jquinn@liverpool.ac.uk; Tel: (+44) 151 794 5498.

## Abstract

### Background

The hominid-specific non-LTR retrotransposon termed SINE VNTR Alu (SVA) is the youngest of the transposable elements in the human genome. The propagation of the most ancient SVA type A took place about 13.5 Myrs ago, and the youngest SVA types appeared in the human genome after the chimpanzee divergence. There are about 2700 insertions of the SVAs in human genome that are responsible for gene structural polymorphisms and modulation of gene expression, however functional and evolutionary significance of these insertions is not clear.

### Results

Functional enrichment analysis of genes associated with SVA insertions demonstrated their link to multiple ontological categories attributed to brain function and the disorders. Moreover, SVA types that expanded their presence in the human genome at different stages of hominoid life history were also associated with progressively evolving behavioural features that indicated a potential impact of SVA propagation on a cognitive ability of a modern human. The SVA-associated genes were highly cross-linked in functional networks suggesting an accumulative impact of functional alterations potentially caused by SVA insertions. The analysis of functional networks also pointed to a relevance of SVA-associated genes to a hormonal and immune modulation of behaviour and a crosstalk between behaviour and reproduction, that would effects inheritable propagation of SVAs and their accumulation in the genome. The presence of functional positive and negative feedbacks in the SVA-associated functional network is discussed in relation to a role of suggested waves of SVA and potentially other mobile elements propagation in evolution of human cognitive features.

### Conclusion

Our analysis suggests a potential role of SVAs in evolution of human CNS and especially emergence of functional trends relevant to social and parental behaviour. It also supports models which explain in part how brain function can be modulated by both the immune and reproductive systems based on the gene expression patterns and gene pathways potentially altered by SVA insertions.

**Background**

Genetic studies have been successful at determining the changes that alter pathways involved in evolution. Transposable elements (TEs), despite long being thought of as 'junk' DNA, have impacted the human genome during its evolution through a variety of mechanisms causing structural variation such as exon disruption, generation of deletions and integration sites, 3' or 5' transduction, non-allelic homologous recombination and exonisation (Beck, et al. 2011; Goodier and Kazazian 2008; Muotri, et al. 2007). Furthermore, TE insertions can supply the cell with novel splice sites, polyadenylation signals, promoters and regulatory domains, epigenetic marks and secondary structures that can reorganise gene expression and build new transcription modules which could underpin the process of evolution (Beck, et al. 2011; Hancks, et al. 2009; Hancks and Kazazian 2010; Piriyapongsa, et al. 2007; Quinn and Bubb 2014; Savage, et al. 2013; Savage, et al. 2014) but also cause disease (Kaer and Speek 2013). Estimates have suggested over 1.5 million TEs in the human genome (Cordaux and Batzer 2009). Non-LTR retrotransposons constitute ~34% of the human genome (Beck, et al. 2011), and a subset of these elements have retained the ability to be mobilised within the genome generating new insertions. They have the potential to cause both "de-novo" germline mutations and somatic mutations which are implicated in disease progression, particularly in cancer and the CNS (Erwin, et al. 2014; Reilly, et al. 2013) .

More than 10,000 TE insertions occurred in the human genome since human-chimpanzee divergence (Mills, et al. 2006). The large number of such TEs in the genome makes an analysis of their specific contribution to evolution very difficult. However, to address this contribution we have focused on building networks of pathways based on the proximity of genes to the integration sites of the hominoid-specific retrotransposons termed SINE–VNTR–Alu (SVA) elements. SVA elements are the youngest of the retrotransposons, and there are approximately 2700 of these elements in the human genome (Cordaux and Batzer 2009; Savage, et al. 2013; Wang, et al. 2005). They stand out from the group of human non-LTR retrotransposons due to their composite structure. Starting at the 5′-end, a full-length SVA element is composed of a (CCCTCT) hexamer repeat region, an *Alu*-like region consisting of three antisense *Alu* fragments adjacent to an additional sequence of unknown origin, at least one variable number of tandem repeats (VNTR) region (Ostertag, et al. 2003), and a short interspersed element of retroviral origin (SINE-R) (Ono 1986). A poly(A) tail is positioned downstream of the predicted conserved polyadenylation signal AATAAA (Ostertag, et al. 2003). Others and we have demonstrated that these SVA domains, in addition to mobilisation, have the properties of transcriptional regulators of gene expression both *in vivo* and *in vitro* (Savage, et al. 2013; Savage, et al. 2014; Zabolotneva, et al. 2012) , and that a significant number of SVAs are within 10kb of the major transcriptional start site of many genes (Savage, et al. 2013). Therefore, SVA insertions established in the hominoid lineage could be responsible for altering the transcriptome in a developmental, tissue specific or stimulus inducible manner. The relatively low number of SVAs established in the genome allows for a model to determine the gene pathways likely to be affected by the SVA integrations.

SVAs are divided into subtypes A to F and F1, and their age was estimated at 13.56 million years (Myrs) for the oldest subtype (A) and 3.18 Myrs for the youngest subtype (F). Subtype D is by far the largest and encompasses 44% of all SVAs in the human genome. The most recent F1 subtype is the smallest group at 3%. Subtypes E, F and F1 are human-specific and correspond to the period since the human-chimpanzee divergence ~6 million years ago (Mills, et al. 2006). Implication of SVAs in evolution of hominoid lineages leaves no doubts. After the human-chimpanzee divergence, the SVAs continued to expand within the chimpanzee genome creating a subtype unique to chimpanzees called SVA PtA (Wang, et al. 2005). More recently, the family of gibbon-specific LAVA retrotransposons, derived from an SVA subtype A element, has been implicated in the molecular mechanism underpinning genome plasticity of the gibbon lineage (Carbone, et al. 2014). It was reported that LAVA-induced premature transcriptional termination of chromosome segregation genes caused the high rate of chromosomal rearrangements experienced by the gibbon lineage since it diverged from the other apes about 17 Myrs. An analysis of the human and chimpanzee genomes revealed that 46537bp have been deleted from the human genome through the processes of SVA insertion-mediated deletions and SVA recombination associated deletions (Lee, et al. 2012).

In this communication, we focus on the SVAs that are part of the human genome and address the functional relevance of human SVA-associated genes to neurological and cognitive processes to demonstrate SVA insertion and subtype appearance in evolution with the correlation of behavioural traits. Our data also support evidence that SVA insertions can have an impact in normal and pathological brain functioning (Richardson, et al. 2014; Upton, et al. 2015).

## Methods

### Generation of the list of genes associated with SVA insertions

Genomic coordinates of all SVA loci in the human genome (Hg19 sequence) were extracted from the UCSC genome browser (http://genome.ucsc.edu/index.html). This included many SVA sequences that were fragmented in the Repeat Masker track; therefore this list was manually annotated to generate a list of coordinates of complete SVA sequences resulting in a total of 2676 of these elements encompassing the seven subtypes (Savage, et al. 2013). The coordinates of all known genes and their transcripts were extracted from the UCSC genome browser, and Galaxy software (http://galaxyproject.org/) was used to generate coordinates of the 10-kb genomic regions flanking all known transcripts. Finally, the SVA loci were intersected with the three lists of genomic coordinates (all known genes, 10kb upstream and 10kb downstream of known genes). Duplicates were removed from each list individually. For functional enrichment analysis, the defined SVA coordinates have been used to produce a shorter list of GRch38 genes directly mapped via BIomart Martview service (www.biomart.org/biomart/martview/).

**Ingenuity Pathway Analysis (IPA)**

Ingenuity Pathway Analysis (IPA) software (Ingenuity Systems, Inc.) was used to investigate biological pathways and networks as well as disease functions associated with the data set (Gietzen 2010; IPA 2013; Jiménez-Marín, et al. 2009). The IPA database is composed of comprehensive information on genes and pathway ontologies. The database contains approximately 200 canonical pathways as well as 27 higher-order disease and function categories for 'Core' functional enrichment analysis. Right-tailed Fishers Exact statistical tests was used to calculate whether the likelihood of associations between a set of focus genes and a category was due to a random chance. This enabled evaluation of enriched functions in a pathway or higher order ontological categories and provided a hypergeometric distribution based network score and p-Values conveyed as the –log (Fisher's Exact Test). The tool enabled production of graphical networks that illustrated mapped genes as shadowed objects. Un-shadowed objects were added as connectors within a network, identified by the software. The behavioural categories used by IPA are largely originate from experiments on rodents and may seem as irrelevant to human. However, parallels between mice and human behavioural features are widely used in neurobiology, and we try to provide these parallelisms/explanations of relevance to human were appropriate.

**Meta-analysis of gene expression.**

'Genevestigator' (Grennan 2006; Hruz, et al. 2008; Zimmermann, et al. 2005) software was used to perform a meta-analysis of gene expression in different tissues and areas of the human brain. For each gene/tissue an average level of expression is determined via automated cross-analysis of normalised published Agilent microarray data, filtered for statistically significant values and stored in the Genevestigator database. The highest and lowest detected average expression levels define a 100-unit scale that is used for presentation of each gene/tissue related average expression value.

**Results**

## Relevance of SVA-associated genes for functional categories of the neural system

Having coordinates of SVA insertions in human genome we aimed to identify if there are any functional biases in the associated gene set. We have confined our functional enrichment analysis to those genes, which imbed an SVA insertion. Though without significant enrichment, the top (from 2nd to 6th positions) ranked 'Canonical pathways' were uniformly relevant to neuronal functions including: 'Synaptic Long Term Potentiation', 'Synaptic Long Term Depression', 'Axonal Guidance Signalling', 'Neuropathic Pain Signalling in Dorsal Horn Neurons' and 'CREB Signalling in Neurons' (Table S3A). The gene functions contributing to these canonical pathways were associated with the top five 'Functions and Diseases' categories , where 'Guidance of axons' and 'Synaptic processes' were significantly enriched (Table S3B). The presence of multiple top ranked neural system links was indicative of a potential impact of SVA insertions on neural system function at different stages of human cognitive and behavioural evolution. Moreover, significant enrichment of genes in a particular pathway or functional category could mean evolutionary functional stratification of the SVA insertions via positive selection of the relevant phenotypes. The results of our analysis overall do not suggest that neural system was the main target of SVA attack, as the significant enrichment was observed only for two relevant functional categories . However, it was still interesting to see if we could uncover any specific functional trends of potential SVA impact on human cognitive function.

## Analysis of the functional enrichment of SVA-associated genes with neuronal functions.

To focus on potential significance of SVA insertions in function and evolution of neural system only genes representing relevant to CNS categories ('Behaviour', 'Nervous System Development & Function', 'Neurological Diseases' and 'Psychological Diseases' ) were selected (Tables S1 and S2) and ontologically classified by means of IPA (Table S4). The analysis of their biological functions showed a significant enrichment in categories of 'Morphology of nervous system' (-log(p-value)=17.9) (forebrain, and telencephalon, particularly), 'Development of central neural system' (-log(p-value)=16.3) (forebrain, and telencephalon, particularly), 'Formation of cellular protrusions' (-log(p-value)=12.2) and multiple categories relevant to axon growth and synaptic processes (-log(p-value)>5). From all the mapped Behavioural categories, 'Learning' and 'Social behaviour' were the most enriched (-log(p-value)>4.6) (Table S4), and 'Nest building behaviour', 'Nursing', 'Sexual behaviour', 'Sexual receptivity of female organism', 'Grooming', 'Emotional behaviour', 'Aggressive behaviour' , 'Vocalisation' and 'Walking' significantly enriched in the SVA B and/or SVA D associated sets of genes. 'Social behaviour' was the only significantly enriched category attributed to the gene sets associated with the youngest, F and F1 SVAs. Many categories across the sets were mapped by one or two genes (Table S5) and were not considered as significant. All the defined behavioural categories were presented against the timeline of occurrence of insertions of the different SVA subtypes in the human genome as shown on Figure 1.

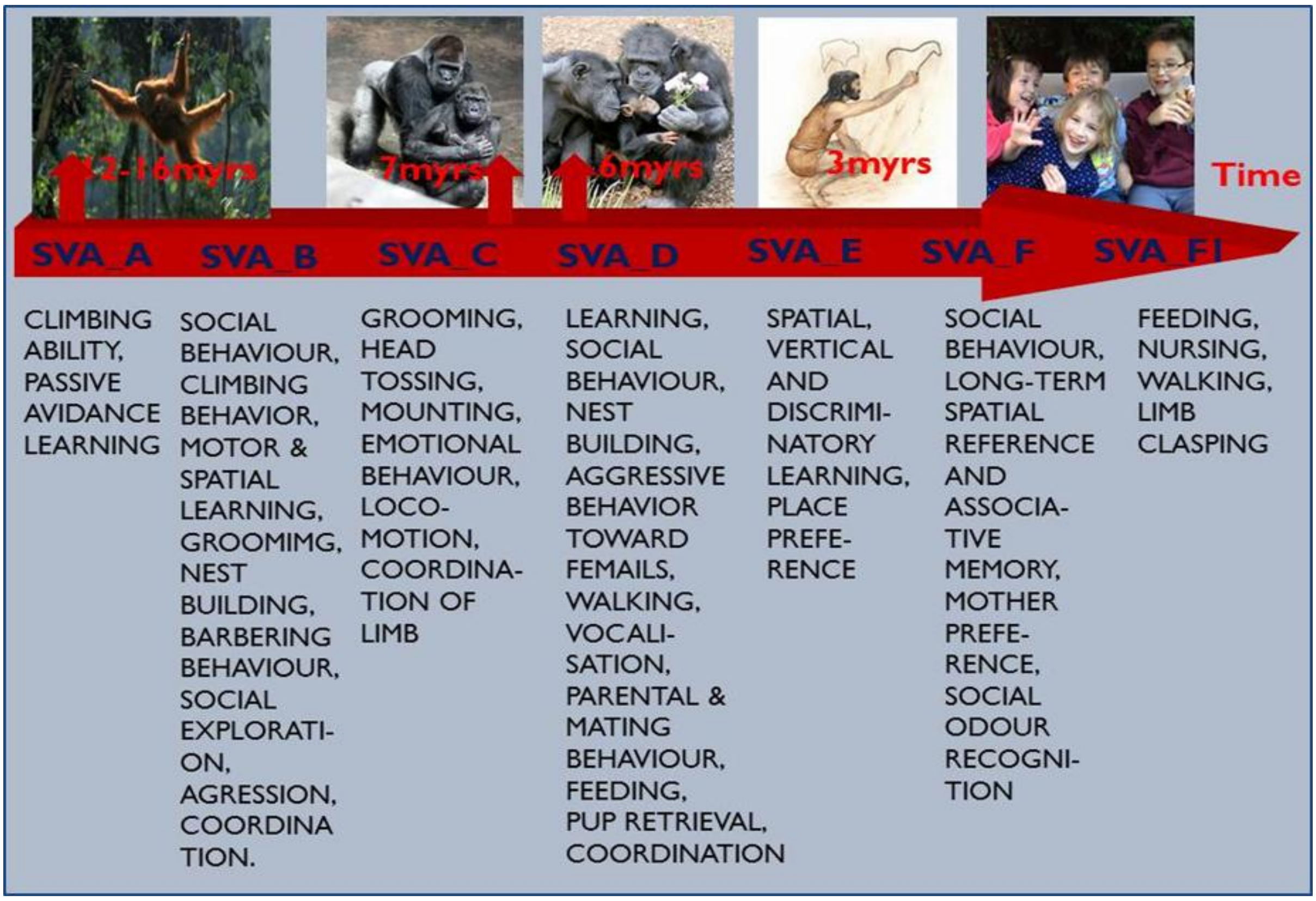

Fig.1. Behavioural categories attributed to genes associated with SVAs at different time points during evolution.

Our results suggest a functional relevance of SVA-mediated modification of individual genes, which could be attributed to the evolutionary behavioural changes. Some categories used by IPA ( such as nest building) and consequently, functional meaning of potential modification of gene activity due to SVA insertion may be not obvious at first glance. Nevertheless, without going through all the genes in this list, we have outlined below using few examples how SVAs could be involved in evolution. It can be seen that the gene ATG7 (Autophagy related 7 homolog), that contains an SVA F1 insertion, is associated in the database with the 'Walking' functional category (Table S5,), supporting the hypothesis that younger SVA subtypes could affect the particular locomotor patterns of pre-modern human favouring emergence of the new characteristics. The walking-category-associated group of genes (AGTPBP1 (ATP/GTP binding protein 1), CACNB4 (Calcium channel, voltage-dependent, B4 subunit), FXN (Frataxin), SCN8A (Sodium channel, voltage gated, type VIII alpha subunit) also contain SVA D insertions (Table S5) which potentially reflects their involvement in development of erect walking, characteristic of a chimpanzee-hominid phylogenetic branch (Figure 1). Conversely, the functions of climbing activity-associated genes KCNJ6 (Potassium inwardly-rectifying channel, J6) and HTT (Huntingtin) could be modulated or suppressed by SVA A and B insertions very early in hominoid evolution (Table S5). Alteration of the functionality of these genes overall could have had a strong impact on changing from climbing to the walking moving mode. Mother preference associated gene OMP (Olfactory marker protein) was present in the SVA F subtype group as were the nursing and feeding associated genes ACOT11 (Acyl-CoA thioesterase) and MKL1 (Megakaryoblastic leukaemia (translocation) (Table S5). These latter genes may be proposed for a role in the progression of maternal behaviour and neotenic features in human evolution (Figure 1). Behavioural characteristics encoded by genes containing younger SVA elements are generally of a more executive nature (Table S5): for instance, CREB1 (cAMP responsive element binding protein 1), associated with an SVA E insertion, is involved in both vertical and spatial learning and place preference, while PSEN1 (Presenilin-1) is associated with an SVA F element and plays a role in social order recognition, memory and social behaviour.

Several genes in our list are associated with multiple categories, e.g. the genes HTT (SVA B subtype group) or BRCA1 (Breast cancer type 1 susceptibility protein) (SVA F subtype group) could have been especially significant at particular evolutionary stages whilst having largely pleiotropic effects (Figure 2). Related functional networks enriched in genes with SVA insertions and belonging to neuronal/behavioural categories (Figure 2A, B) showed that genes harbouring different SVA subtypes interact and the changes in gene expression caused by SVA insertions may be synergistic with an accumulative or compensatory impact on particular phenotypes. Interestingly, 13 functions have been found to be associated with multiple different (at least 3) SVA subtypes (Table 1), from which six genes occurred to be connected in one functional network (Figure 3). The network connectivity of this gene set (Figure 3) demonstrates direct links from the genes BCYRN1 and MAP4 to X-chromosome fragility susceptibility gene FMR1, and functional association with the PGR (Progesterone receptor) and ZNF640 (Zink finger 640) via mir-548 family of regulatory RNAs. Interestingly, 24 additional genes attributed to X-linked mental retardation were present in our SVA-associated data set (Table 2). Categories relevant to X-linked mental retardation were significantly enriched in several gene sets (Figure 4); it was also shown for ‘Seizure’, ‘Movement’ and ‘Mood ‘, Alzheimer, Parkinson CNS disorders and Schizophrenia in association with different SVA types (Figure 4).

**Table 1**. Genes associated with 3 and more SVA insertions. The presence of a particular SVA type in a gene or in 10kB gene proximity is indicated by '+'.

| Gene | Function<br>SVA: | A | B | C | D | E | F | F1 |
|---|---|---|---|---|---|---|---|---|
| **BCYRN1** | Brain cytoplasmic RNA 1 | + | + | | + | | + | |
| **DLEC1** | Deleted in lung and esophageal cancer 1 | | + | + | | | | + |
| **EIF4G3** | Eukaryotic translation initiation factor 4 gamma, 3 | | | | + | + | + | |
| **TAF1** | TAF1 RNA polymerase II, (TBP)-associated factor | + | | | + | | | |
| **SIMC1** | SUMO-interacting motifs containing 1 | | | | + | | + | |
| **CCDC109B** | Coiled-coil domain containing 109B | + | + | | + | | | |
| **WDR70** | WD repeat domain 70 | + | | + | + | | | |
| **DST** | Dystonin | | + | | + | + | | |
| **UBR1** | Ubiquitin protein ligase E3 component n-recognin 1 | | + | | + | + | | |
| **KDM4C** | Lysine (K)-specific demethylase 4C | | | + | + | | + | |
| **MAP4** | Microtubule-associated protein 4 | | | + | + | | + | |
| **mir-548** | MicroRNA 548c | | | | + | + | + | |
| **PTPN9** | Protein tyrosine phosphatase, non-receptor type 9 | | | + | + | | | + |

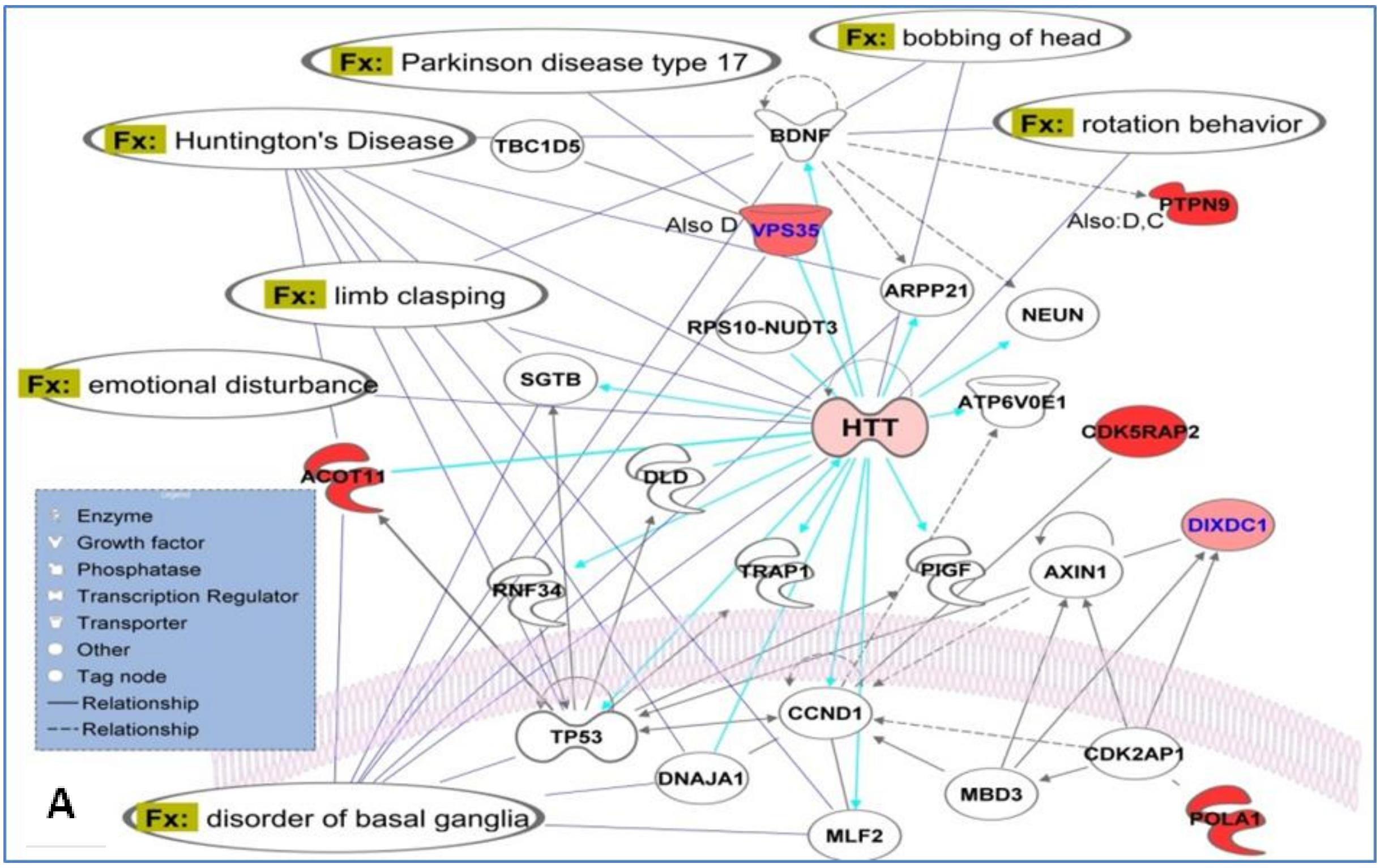

Fig.2. Functional networks reconstructed from SVA-associated genes classified in neuronal and behavioural categories. The initial networks reconstructed by IPA from uploaded dataset (coloured blocks) have been refined by deletion of automatically added low-connected nodes. Fx-labels correspond to gene-attributed functions and diseases. Solid lines reflect experimentally validated functional interactions, and dashed lines indicate protein binding. A) A network of genes

associated with HTT (Huntingtin), who's connections are highlighted in blue. The colour intensities of blocks correspond to different SVA-subtypes increasing in the order: B, C, D, F1. Presence of multiple SVA subtypes is indicated in associated textboxes. The majority of interacting genes has intragenic SVA insertions; presence of upstream or downstream insertions is indicated by blue lettering.

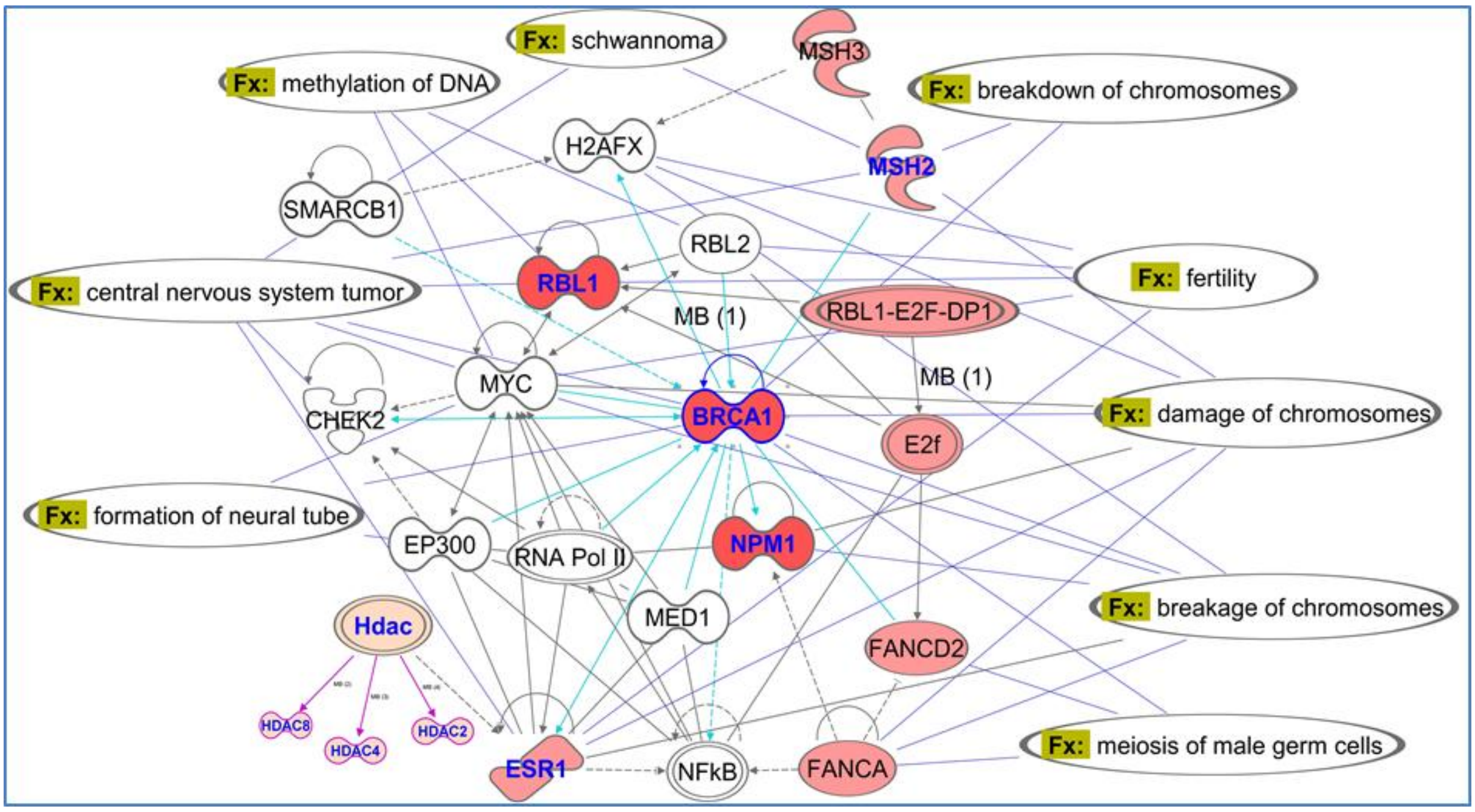

B) A top-ranked network with assigned relevance to DNA recombination and repair associated with BRCA1 (Breast cancer associated 1), who's connections are highlighted in blue. The colour intensities of blocks correspond to types of intragenic SVA insertions increasing in the order: B, D, F. Blue lettering indicates the genes associated with neuronal and behavioural categories. ACOT11-Acyl-CoA thioesterase 11, CDK5RAP2-CDK5 regulatory subunit associated protein 2, DIXDC1-DIX domain containing 1, E2F-Elongation factor 2, ESR1-Estrogen receptor1, FANCA-Fanconi anemia protein A, FANCD2-Fanconi anemia protein D2, HdaC-Histone deacetylase, MSH2-MutS homolog 2, NPM1-Nucleophosmin, POLA1-Polymerase (DNA directed), alpha 1, catalytic subunit, RBL1-Retinoblastoma-like 1, VPS35-Vacuolar protein sorting 35 homolog .

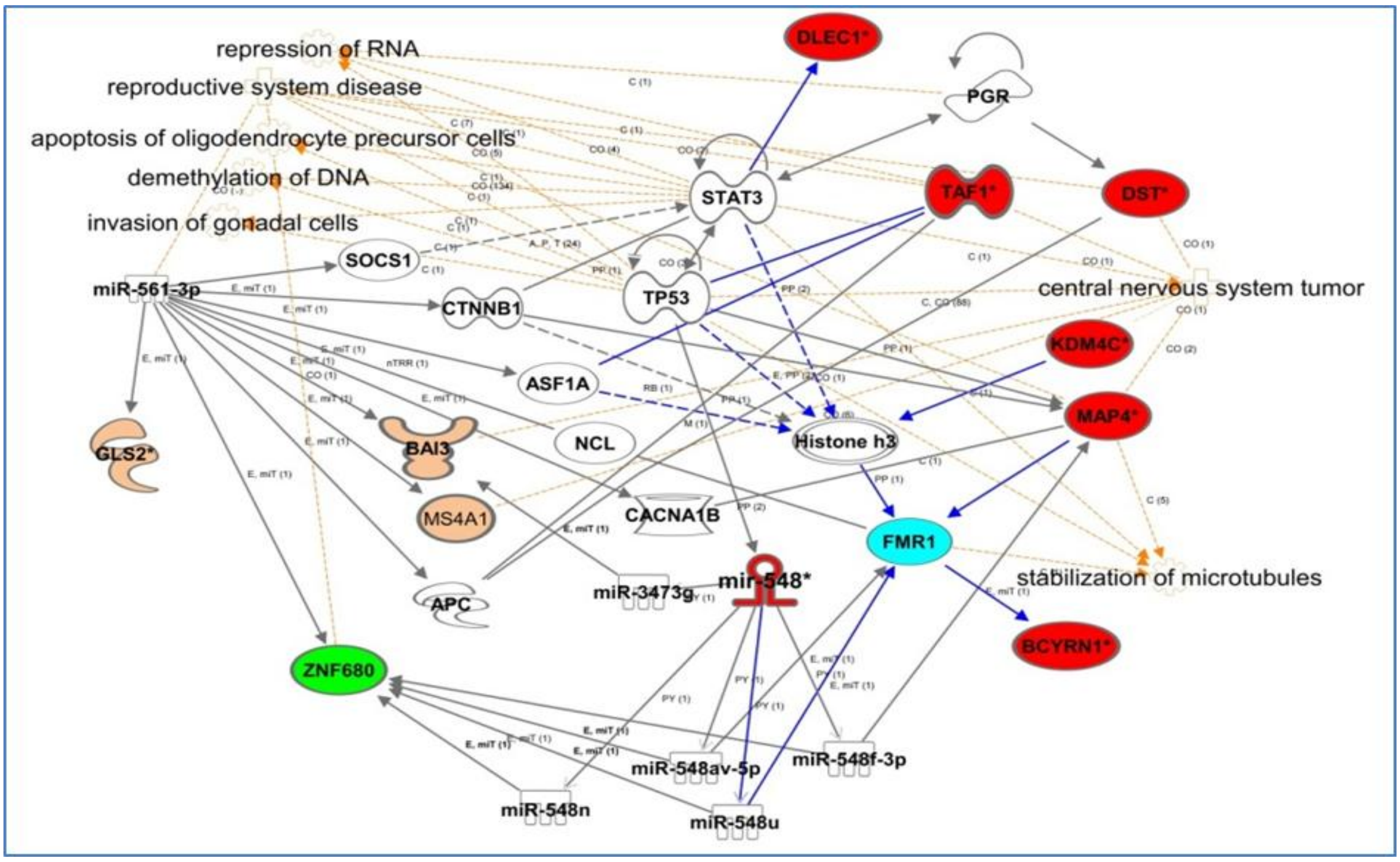

Fig.3. A network reconstructed from genes (red objects) with three and more different intragenic SVA insertions. Associated genes containing less than three SVA insertions are presented in pink and linking genes- in white . X-chromosome fragility gene FMR1 and the KRAB-family ZNF680 are highlighted in blue and green, respectively. Linked text labels (connected by orange lines) correspond to gene-attributed functions and diseases. Solid lines reflect experimentally validated functional interactions and dashed lines correspond to protein binding. Shortest paths from SVA-associated genes to FMR1 are highlighted in blue. BCYRN1 -Brain cytoplasmic RNA 1, lncRNA,TAF1 -TATA box binding protein, DLEC1-Deleted in lung and oesophageal cancer 1, EIF4G3-Eukaryotic translation initiation factor 4 gamma, Map4-Microtubule-associated protein, mir-548 – micro RNA.

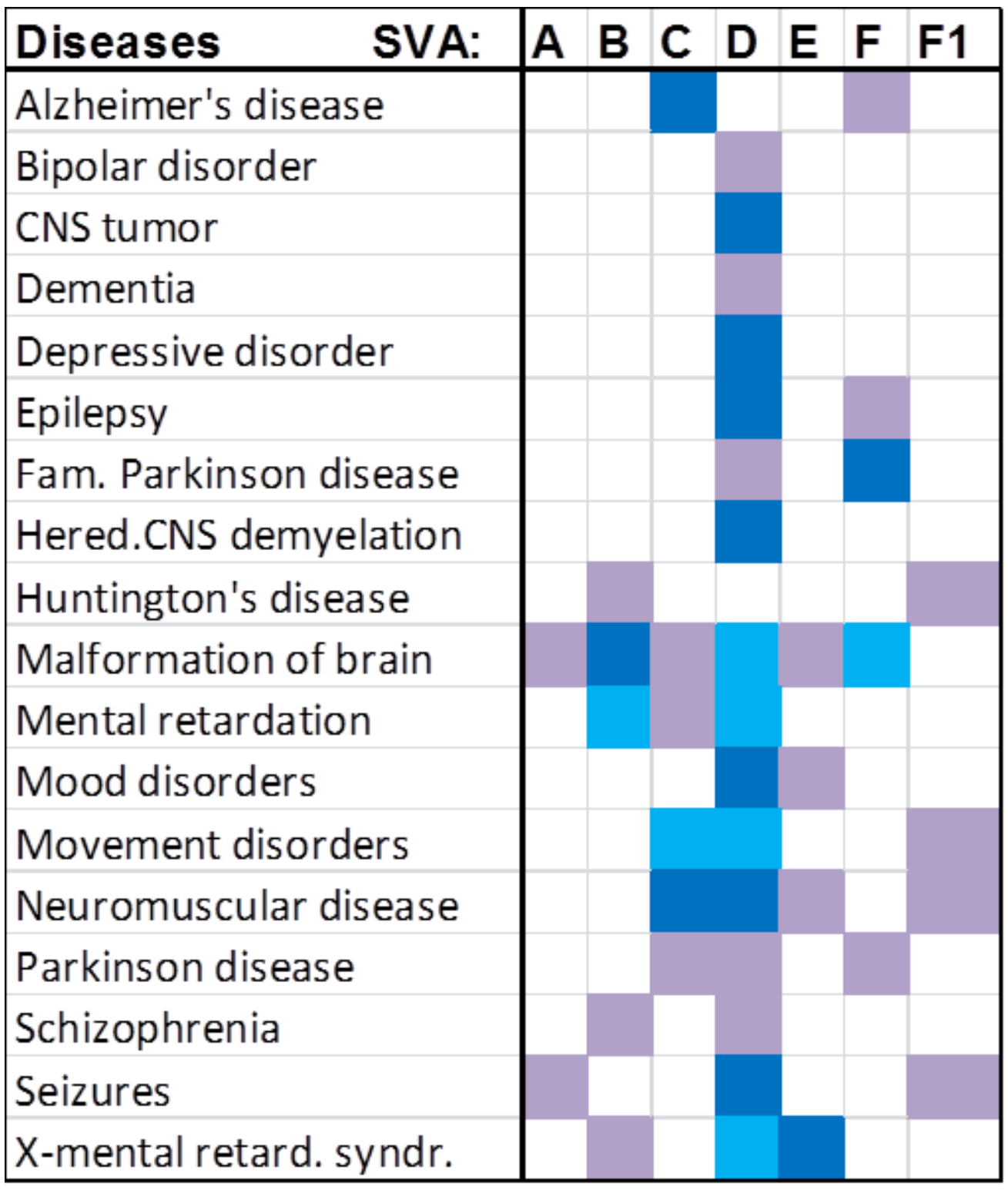

Fiig.4. Link of genes associated with SVA insertions to neural and psychiatric diseases. Colours correspond to particular types of SVA (columns) in genes from certain disease categories (rows). A significant enrichment (p-value<0.05) of the neuronal/behavioural gene dataset in the a category is highlighted in blue, light blue corresponds to a higher ( p<0.01) significance of the functional enrichment.

**<u>Meta-analysis of anatomical relevance of SVA-associated behavioural gene expression.</u>**

The forebrain and hidbrain were confirmed to be the sites of high expression of the majority of the genes attributed to behavioural phenotypes. High expression levels of the majority of genes associated with all 7 SVA subtypes were also found outside of the CNS. Potential positive selection of SVA-associated genes could be attributed among other factors to their direct effect on an organism's reproduction. Nearly all genes associated with members of the SVA A and B subtypes were highly expressed in placental chorion. All genes were expressed in testis at above average levels. A number of the behavioural SVA-associated genes exhibited in oocytes the highest expression levels (TIMD4, OXT, KALRN, MKL1, ATG7, cyp7B1) with noticed enrichment of such genes in SVA F1 and SVA F associated. Some genes whose expression results in a validated, strong 'neural' phenotype, such as GABBR2 (SVA A) and CREB1 (SVA E), are also highly expressed in oocytes. Interestingly, the germline cells are reported to be sites of increased mobilisation of endogenous retrotransposons in general (Levin and Moran 2011), and attribution of SVAs to genes expressed in these tissue might therefore be relevant to the SVA inheritable propagation. Analysis of attribution of the gene set with neuronal and behavioural functions in phenotypic reproductive categories (IPA analysis) also highlighted their link to: spermatogenesis (sperm motility (ADCY10), atrophy of testis and seminiferous tubules (HTT), in particular), infertility in both genders (STAR, ABCA1, MKKS, NPEPPS), maturation (LFNG) and fertilisation (CLIC4) of oocytes. Several genes, such as CYp19A1, ESR1, ABL2, CREB1, NPEPPS and SHANK1 also have an association in the IPA knowledgebase with behavioural aspects of reproduction: sexual receptivity of female organism, mating behaviour, intromission, mounting and pup retrieval.

High expression levels of the SVA-associated genes, in particular of those with the youngest SVA subtypes (APL2, OMP, PSEN1, and ATG7, for instance) and few with older subtipes (Hpp) were also observed in the immune system. This was of interest given the current attention to the immune–CNS axis and how interactions between the immune system and CNS may not only regulate normal physiology but are also be associated with mental health issues. HTT was the only gene of the gene group associated with SVA A or SVA B elements that was strongly expressed in cells of the immune system.

## Discussion

Retrotransposons have been suggested to be major drivers of genome evolution, both through structural variations such as insertional mutagenesis and modulation of the transcriptome. For the latter, retrotransposons are a source of regulatory elements providing promoters (sense and antisense), binding sites for transcription factors and polyadenylation signals that could affect gene expression (Rebollo, et al. 2012). SVAs can also induce an alternative splicing and exon skipping which can result in the generation of alternative transcripts of a gene as documented by disease causing insertions (Hancks and Kazazian 2012; Kaer and Speek 2013; Nakamura, et al. 2015). Furthermore, the SVA encoded SINE-R module contains human endogenous retrovirus HERV-K10 LTR sequences that are suggestively involved in the regulation of gene expression (Fairbanks, et al. 2012; Xing, et al. 2006); and at least in case of SVA D, they indeed supported reporter gene expression *in vivo* and *in vitro* (Savage, et al. 2013; Savage, et al. 2014). Thus, SVA insertions may generate new interactions between what would otherwise be distinct pathways (Quinn and Bubb 2014).

SVAs share many DNA sequence homologies across subtypes and, as a result, could respond to related or similar stimuli giving a concerted gene response to the environment or challenge. We have therefore focused on pathways involved in CNS function and predicted to be modulated by human genes encompassing or adjacent to SVA insertions. Our data has established that more primitive behavioural characteristics are prominent in the genes associated with the older SVA subtypes A, B and C. Results revealed a striking correlation between the timing of SVA insertions and functional significance of these associated genes at particular stages of hominoid evolution. For example, genes associated with SVA A and B insertions (which expanded contemporary to the divergence of the orang-utan and the great apes (Wang, et al. 2005)), influence climbing ability. In contrast, genes containing SVA F insertions (which expanded after the human and chimpanzee divergence(Wang, et al. 2005)) were found to affect more advanced motor skills such as walking (Figure 1). Similarly, the analysis revealed that complex behavioural characteristics (such as discriminatory learning, sexual, social and especially maternity care) correlated with the more recent SVA insertions (Figure 1). Meta-analysis of gene expression data revealed that genes associated with older SVA insertions are more strongly expressed in structures of the brain that evolved earlier, such as the pyramidal regions. It is also noteworthy that SVA F and F1 insertions were strongly associated with genes involved in the immune system which is considered a major modulator of CNS function, and in neuropsychiatric disorders such as autism, schizophrenia and depression [TC Network (2015)]. Interestingly, increased brain size in mammals was also shown to be also associated with size variations in immune system-involved gene families (Castillo-Morales, et al. 2014). Late age-related phenotypes of SVA insertions could also play a role in reduction of the reproductive age and general neotenisation of the human population.

SVA insertions could also have an influence on human gender-specific brain function via sensing hormonal backgrounds. The BRCA1 gene (Figure 2B), which is associated with an SVA F insertion, was shown to be in a functional cross talk with the oestrogen receptor ESR1 (Gorrini, et al. 2014; Kang, et al. 2012). It is also known to have a strong role in a control of brain size (Pao, et al. 2014; Pulvers and Huttner 2009), especially in regions responsible for learning, memory, muscle control and balance. Interestingly, progesterone-dependent transcription factor, PGR, which binding sites were shown to be enriched in ancient mammalian TEs (Lynch, et al. 2015), was also associated with a network of genes with multiple intragenic SVA insertions (Figure 3).

There is an intriguing correlation in expression of many SVA-associated genes in brain and testis, and a very strong expression of some of these genes with behavioural phenotypes in oocytes. It may be relevant to regulation of the genes via putative alternative promoters (PAPs)(Kimura, et al. 2006) especially enriched in those tissues and potentially associated with SVA elements. Moderate expression of the majority of behavioural- SVA-associated genes in male germ cells (in contrast to only few, though strongly, expressed in oocyte) could make this cell type a particular source of mobilized SVAs and also a place for their inheritable *trans*-mobilisation. It would be supported by a strong expression in this niche (as well as in an early embryo) of retroviruses and L1 retrotransposons as the required providers of the reverse transcriptase (Han, et al. 2004; Levin and Moran 2011; Zamudio and Bourc'his 2010). The observed correlation means that certain behavioural trends caused by SVA insertions could be genetically linked to a higher rate of an inheritable SVA *trans*-mobilisation. Being subjected to positive or negative selection at particular stages of hominoid evolution those behavioural trends therefore could be associated with waves of inheritable SVA propagation in the genome. Such activated *trans*-mobilisation of SVAs in its turn could lead to increased genome instability and chromosome fragility affecting also the rates of somatically-driven CNS disorders. It would finally suppress a positive selection of genotypes favouring such inheritable trends. The presence of the hypothesised feedbacks is supported by a number of our observations. DNA recombination and repair were amongst the top categories in functional groupings enrichment analysis of all SVA associated functions (Figure 2B, Table S3), and X-fragility-relevant genes populated the list of behavioural SVA-associated ones (Figure 3, Table 2), which may reflect on the mechanisms of how SVAs can affect their propagating properties. This is also consistent with reported genome instability caused by mobile elements (Kim and Hahn 2011). Genes associated with both neuronal and reproductive phenotypes as well as with SVA *trans*-mobilisation properties (for instance, BRCA1 or the KRAB proteins (ZNF680 & ZNF91/93) (Jacobs, et al. 2014; Thomas and Schneider 2011) could be in the first line of the SVA-driven evolutionary changes in the hominoid lineage (Figure 4). The network reconstructed from those genes, that are especially enriched in SVA insertions (Figure 3), highlights interacting negative (KRABs, miRNAS) and positive (BCYRN1, miRNA) pathways, which could be extended to gene silencing and RNA interference by the piwiRNA pathway (Figure 5) (Bao and Yan 2012; Kapusta, et al. 2013; Zamudio and Bourc'his 2010) in regulation of SVA propagation.

SVAs are present in genes, which have been linked to different psychiatric conditions ranging from intellectual disability (mental retardation) to epilepsy and bipolar disorder (Figure 4). Human specific SVA insertions are known to be polymorphic for their presence or absence (Bennett, et al. 2004), and with continued retrotransposition there are likely to be many private insertions within a population. The polymorphic nature of SVA elements is also exhibited in terms of the number of repeats within their central VNTR region and hexamer domains (Savage, et al. 2013). VNTRs, as individual elements, are associated with the differential expression of genes involved in human behaviour such as those found with the genes encoding monoamine oxidase A, serotonin and dopamine transporters (Hill, et al. 2013; Lovejoy, et al. 2003; Michelhaugh, et al. 2001) and constitute risk factors for a variety of behavioural disorders and psychiatric diseases dependent on the copy number of the repeat present(Savage, et al. 2013)(Savage, et al. 2013)( (Ali, et al. 2010; Breen, et al. 2008; Galindo, et al. 2011; Hill, et al. 2013; Klenova, et al. 2004; Paredes, et al. 2012; Pickles, et al. 2013; Roberts, et al. 2007; Vasiliou, et al. 2012). Furthermore, the VNTR domain plays a major role in the *trans*-mobilisation of SVA elements by the L1-encoded protein machinery (Han, et al. 2004; Hancks and Kazazian 2010; Raiz, et al. 2012).

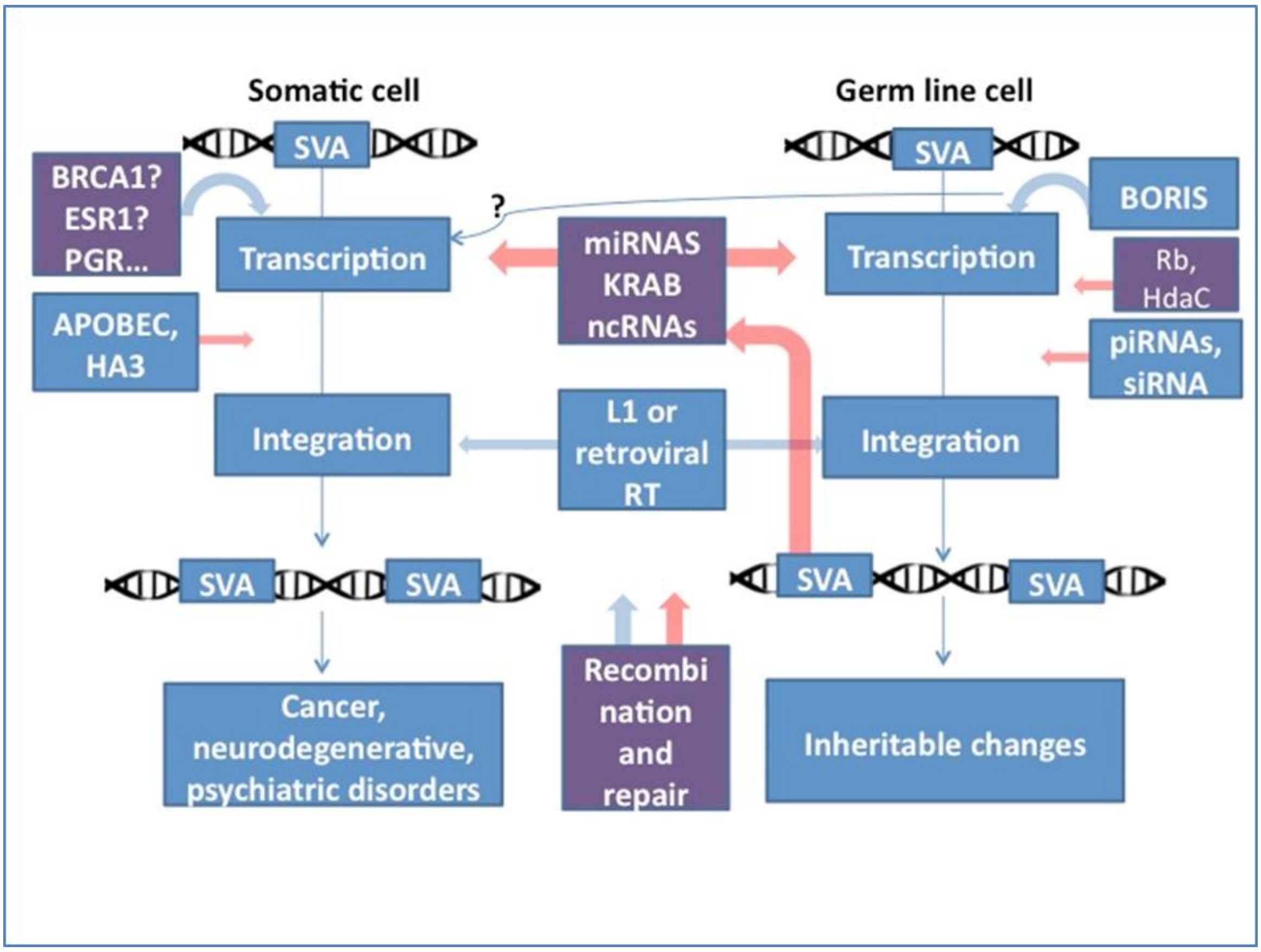

Fig.5. Schematic presentation of regulatory and evolutionary functional interactions involved in SVA trans-mobilisation. Bold arrows correspond to inhibitory (blue) or activatory (red) effects. Functions targeted by SVA insertions are highlighted in purple. RT-reverse transcriptase, Rb-Retinoblastoma protein, HDAC-histone deacetylase, piRNA-piwiRNA, APOBEC-Apolipoprotein B mRNA editing enzyme, catalytic polypeptide-like. Regulators of L1 expression are not shown.

**Conclusions**

Our systems analysis suggests an impact of SVA retrotransposition on evolution and performance of human cognitive functions. It has revealed that insertions of members of the older SVA subtypes A-C occurred in genes involved in more primitive characteristics whereas younger SVA D-F1 insertions were present in genes linked to more sophisticated human-specific traits. SVA insertions were also found to be enriched in genes involved in networks and pathways relevant to neuronal function and CNS disorders. The presence of SVAs in multiple genes within a network may allow for a concerted response to an environmental stimulus modulated by factors targeting SVA regulatory sequences. Potential positive feedbacks in functional interactions of genes with inserted SVAS would have a larger impact on phenotype compared to if only a single gene within a pathway contained an SVA insertion. Negative feedbacks in these functional networks and co-evolution of factors that limit SVA propagation, from the other side, would play a genome-stabilising role. Taking in consideration genetic polymorphisms in human population associated with SVA insertions and documented somatic mobility of the youngest SVAs in human brain and germ tissues, one can suggest that SVAs also have a strong modulatory impact on recent human behavioural trends and human susceptibilities to psychiatric disorders.

**Availability of Data and Materials.**

Table S2 presents Gene Bank IDs for all the genes mentioned in the paper. Genomic coordinates of all SVA loci in the human genome (Hg19 sequence) were extracted from the UCSC genome browser (http://genome.ucsc.edu/index.html) (Please see Methods for the details).

**List of abbreviations**

CNS- Central Nervous System. IPA-Ingenuity software. All Gene abbreviations are explained in Table S2.

**Competing interests**

There are no any competing interests regarding this manuscript.

**Authors' contributions**

OV conceived of the study, designed and coordinated the systems analysis, and drafted the manuscript. SC took part in the analysis and drafting of the manuscript, AS has performed the retrieval of SVA –associated gene sets and helped to draft the manuscript, GS participated in data analysis and helped in drafting the manuscript, VB participated in retrieval of SVA-associated gene sets and helped in drafting the manuscript, JQ conceived of the study, coordinated SVA-associated gene set retrieval and took part in drafting the manuscript. All authors read and approved the final manuscript.

**Acknowledgements.**

We would like to express our gratitude to Dr. Nathan Jeffery (University of Liverpool) for his very useful comments on evolutionary relevance of the discussed behavioural functions.

**Supplementary material ( attached below)**

Table S1 (Genes associated with SVAs classified to 4 categories relevant to neural system function) and Table S2 (SVA-associated genes with neuronal function and their SVA insertion locations (Upstr.-within 10kb upstream, Downst.-within 10Kb downstream a gene ). The tables provide the lists of SVA-associated genes, Gene Bank IDs and information on a particular type of a gene-associated SVA.

TableS3. Enrichment of SVA-associated gene functions in IPA knowledge base categories. Functional groups directly attributed to neuronal function are highlighted. **A.** Canonical pathways. **B.** Diseases and Functions.

Table S4. Link of SVA-associated genes to neural system related ontological categories. Shading represents an association between genes with particular types of SVAs (columns) with certain categories (rows). An enrichment in a category with (-log(p-value)>2.9) is highlighted in purple and (-log(p-value)> 4.6), in blue.

Table S5. Genes grouped by their associated behavioural categories and the SVA-types (a (SVA_A)-g(SVA_F1)).

**Supplementary materials:**

**Table 1.** Genes associated with SVAs classified to 4 categories relevant to neural system function.

| SVA_A | SVA_B | SVA_C | SVA_D | SVA_E | SVA_F | SVA_F1 |
|---|---|---|---|---|---|---|
| | | | BEHAVIOUR | | | |
| GABBR2 KCNJ6, PLCB4 | CLCN3, HDAC4, HTT,MAPK1, RCAM, NRXN1, OXT, SEC24D, TACR3, TIMD4, UBR1, USF1 | ABCA1, ABHD12, ATF2, FGF12, LRRK2, MKKS, NPEPPS, SLC16A2 | ABL2, AGTPBP1, ATP7A, ATP8A2, CACNB4, CNTNAP2, CYP19A1, ESR1, F2RL1, FTO, FXN, GABRA2, GABRG3, GFRA2, GNAQ, GPHN, IL1RN, LEPR, LRPAP1, MOS, MTOR,NFATC3, NIPBL, NRXN1, OCLN, PARK7, PBX3, PRKG1, SCN8A, SHANK2, SYTL4,TRPV1, UBR1, VAMP7, VEZT | CREB1, OPHN1, UBR1 | APLP2, KALRN, LEF1, MYH10, OMP, PARK2, PDE11A, PSEN1 | ACOT11, ATG7, CYP7B1, MKL1 |
| | | | NERVOUS SYSTEM DEVELOPMENT & FUNCTION | | | |
| ADCY10 AKAP9, CASP6, GABBR2, KCNJ6, PHB, PLCB4, SMO, TRA, ULK4, UNC13B | ARHGAP5, CLCN3, CNTN4, DCLK1, DST, FBXL20, GRID1, HDAC2, HDAC4, HTT, MAPK1, MYO5A, NCOR1, NR1H2, NRCAM, NRXN1, OXT, | ALPL, APBB2, ATF2, HEY2, HYOU1, LAMA2, LRRK2, MKKS, PARG, PTPN9, RORA, SIL1, STAR | ABL2, ADAM19, AGTPBP1, AHI1, ALOX5, ANK2, ARHGAP35, ASPA, ATP7A, ATP8A2, ATRX, ATXN2, BRAF, C2CD3, CACNB4, CAMK1D, CNTNAP2, CRB1, CYP19A1, DDIT3, DIXDC1, EIF2B3, EIF4G3, ELMO1, ERBB2IP, ESR1, F2RL1, FAIM2, FER, FGD4, FTO, FXN, GFRA2, GNAQ, GPHN, IL1RAPL1, IL1RN, JARID2, KIF15, LEPR, LILRB3, LPAR4, LRP4, LRPAP1, MOS, MTOR, MYH9, NCOR1, NFATC3, NIPBL, NRXN1, NRXN2, OCLN, | ADIPOQ, ANXA2, ATP2C1, CASP8, CLIC4, CREB1, DST, LFNG, OPHN1, UBR1, VAV2 | APAF1, APLP2, BRCA1, CLIC4, EIF2B4, ERBB3, FSTL1, KALRN, LEF1, LRP2, LRRC4C, MYH10, NPM1, OMP, | ATG7, CDK5RAP2, CYP7B1, HTR1E, PTPN9, SV2C |
| | PARD3, PDS5B, PRPH2, RPGRIP1L, SMAD2, TACR3, TRPM6, UBR1, UBR2, UTRN | | PARK7, PBX3, PCMT1, PDGFRA, PDS5B, PEX5, POR, PRKG1, PTPN9, RBL1, RGS11, RPGRIP1L, SCN8A, SHANK2, SIL1, SLC17A5, SNTB2, SPTBN1, SYNE2, SYTL4, TACC2, TENM4, TRPV1, TRPV3, UBE4B, UBR1, ULK1, ULK4, UTRN, VAMP7, VEZT, YWHAE | | PARK2, PDE11A, PPARD, PSEN1, RBL1, ROBO2. TENM4, ZEB1 | |
| **NEUROLOGICAL DISEASE** | | | | | | |
| AARS2, ARL13B, BBS9, CASP6, CLN6, DYX1C1, FAF1, GABBR2, KCNJ6, MKS1, PHB, PLCB4, REEP1, RTTN, SMO, TAF1, TRA, ULK4, UNC13B | ACSL4, CACNA2D4, CCM2, CLCN3, DST, FAF1, GRID1, HDAC4, HDAC8, HLA-DRB1, HTT, IGBP1, IGBP1, NR1H2, NRCAM, NRXN1, OXT, PARD3, PARP1, PDS5B, PRPH2, PTPRD, RPGRIP1L, SMAD2, TACR3, TRPM6, UBR1, YARS, ZDHHC15 | ABCA1, ABHD12, APBB2, ARNT, ATF2, CFLAR, DNAJC1, DYSF, FGF12, HYOU1, KDM6A, LAMA2, LOC440040, LPIN1, LRRK2, MFN2, MKKS, PARG, PIGL, PNKD, PTPN9, RAB7A, RORA, SIL1, SLC16A2, SLC25A12, STAR, TFAP2B, VAMP1, WDR62 | ADAM19, ADORA3, AHI1, AIP, ALOX5, ANKRD11, ARHGAP35, ASPA, ATP7A, ATRX, ATXN2,, BBS5, BRAF, BRWD3, C9, CACNB4, CAMTA1, CASK, CFB, CNTNAP2, CPA6, CRB1, CYP19A1, DDIT3, EIF2B3, ELMO1, ESR1, F2RL1, FGD4, FUS, FXN, GABRA, GABRG3, GFRA2, GLRA3, GNAQ,GOLPH3, GPR98, IL1RAPL1, IL1RN, LEPR, LRP4, LRPAP1, MSH2, MTOR, MYH9, NRXN1, OCLN, OTOGL, PARK7, PCMT1, PDE4D PDGFRA, PDS5B, PEX5, PHF8PIGV, PNPT1POR, PTPN9, PUS1, RAB7A, RBL1, RMND1, RPGRIP1L, SCN8A, SIL1, SNTB2, SPRN, SPTBN1, SYTL4, TAF1, TAOK1, TENM4, TRIOBP, TRPV1, UBE4B, UBR1, ULK4, VAMP7, VPS35, XK, YWHAE, ZDHHC9 | ADIPOQ, ANXA2, ATP2C1, BRAT1, CASP8, CREB1, DST, HMBOX1, HUWE1, MYO1B, NBAS, OPHN1, PDE9A, PDXDC2P, RNF135, SNX2, UBR1, VAV2 | ALDH7A1, APAF1, APLP2, BRCA1, CECR2, CTSS, EIF2B4, ERBB3, IFT140, KDM6A, LRP2, MS4A1, NEU1, ORC6, PARK2, PCSK5, PPARD, PSEN1, RANBP2, SERAC1, SLC30A10, SPTLC1, TMC1, VAPB, VPS35 | ACOT11, ATG7, CDK5, RAP2, CYP7B1, POLA1, PTPN9, RAD54L, SLMAP, SV2C |
| **PSYCHOLOGICAL DISEASE** | | | | | | |
| GABBR2 ULK4 | CACNA2D4, HLA-DRB1, HTT, NRXN1, OXT, PTPRD | ABCA1, APBB2, ARNT, CFLAR, DNAJC1, DYSF, FGF12, LOC440040 LRRK2, MFN2, PNKD, STAR, VAMP1 | ADORA3, ALOX5, ANKRD11, ASPA, BRAF, C9, CACNB4, CFB, CRB1, CYP19A1, DDIT3, EIF2B3, ELMO1, ESR1, GABRA2, GABRG3, GOLPH3, IL1RN, LRP4, LRPAP1, MSH2, MTOR, MYH9, NRXN1, PARK7, PCMT1, PDE4D, PEX5, SCN8A,SYTL4, TENM4, TRPV3, UBE4B, UBR1,ULK4, VAMP7,VPS35 | ANXA2, CASP8, CREB1, HUWE1, MYO1B, PDE9A, UBR1 | APLP2, BRCA1, CTSS, PARK2, PSEN1, SLC30A10 SPTLC1, VPS35 | ACOT11, RAD54L, SLMAP |

**Table S2.** SVA-associated genes with neuronal function and their SVA insertion locations (Upstr.-within 10kb upstream, Downst.-within 10Kb downstream a gene).

| | | Gene SVA-insertion location | | | | | | |
|---|---|---|---|---|---|---|---|---|
| Symbol | Entrez Gene Name | Within | Upstr. | Downstr. | Protein location | Molecular function | Entrez Gene ID for Human | Entrez Gene ID for Mouse |
| VAMP1 | Vesicle-Associated Membrane Protein 1 (Synaptobrevin 1) | -- | VAMP1 | -- | Vesicle membrane | other | 6843 | 22317 |
| FUS | Fus fused in sarcoma RNA binding protein | -- | FUS | -- | Cytoplasm | other | 2521 | 233908 |
| AARS2 | alanyl-tRNA synthetase 2, mitochondrial | -- | AARS2 | -- | Cytoplasm | enzyme | 57505 | 224805 |
| ABCA1 | ATP-binding cassette, sub-family A (ABC1), member 1 | ABCA1 | -- | -- | Plasma Membrane | transporter | 19 | 11303 |
| ABHD12 | abhydrolase domain containing 12 | ABHD12 | -- | -- | Other | enzyme | 26090 | 76192 |
| ABL2 | ABL proto-oncogene 2, non-receptor tyrosine kinase | ABL2 | -- | -- | Cytoplasm | kinase | 27 | 11352 |
| ACKR2 | atypical chemokine receptor 2 | -- | -- | CCBP2 | Plasma Membrane | G-protein coupled receptor | 1238 | 59289 |
| ACOT11 | acyl-CoA thioesterase 11 | ACOT11 | -- | -- | Cytoplasm | enzyme | 26027 | 329910 |
| ACSL4 | acyl-CoA synthetase long-chain family member 4 | ACSL4 | -- | -- | Cytoplasm | enzyme | 2182 | 50790 |
| ADAM19 | ADAM metallopeptidase domain 19 | ADAM19 | -- | -- | Plasma Membrane | peptidase | 8728 | 11492 |
| ADCY10 | adenylate cyclase 10 (soluble) | -- | -- | ADCY10 | Cytoplasm | enzyme | 55811 | 271639 |
| ADGRV1 | adhesion G protein-coupled receptor V1 | GPR98 | -- | -- | Plasma Membrane | G-protein coupled receptor | 84059 | 110789 |
| ADIPOQ | adiponectin, C1Q and collagen domain containing | -- | -- | ADIPOQ | Extracellular Space | other | 9370 | 11450 |
| ADORA3 | Adenosine A3 Receptor | ADORA3 | -- | -- | Plasma Membrane | G-protein coupled receptor | 121933 | 562 |
| AGTPBP1 | ATP/GTP binding protein 1 | AGTPBP1 | -- | -- | Nucleus | peptidase | 23287 | 67269 |
| AHI1 | Abelson helper integration site 1 | AHI1 | -- | -- | Cytoplasm | other | 54806 | 52906 |
| AIMP1 | aminoacyl tRNA synthetase complex-interacting multifunctional protein 1 | AIMP1 | -- | -- | Extracellular Space | cytokine | 9255 | 13722 |
| AIP | aryl hydrocarbon receptor interacting protein | -- | AIP | -- | Nucleus | transcription regulator | 9049 | 11632 |
| AKAP9 | A kinase (PRKA) anchor protein 9 | AKAP9 | -- | -- | Other | other | 10142 | 100986 |
| ALDH7A1 | aldehyde dehydrogenase 7 family, member A1 | ALDH7A1 | -- | -- | Cytoplasm | enzyme | 501 | 110695 |
| ALOX5 | arachidonate 5-lipoxygenase | ALOX5 | -- | -- | Cytoplasm | enzyme | 240 | 11689 |
| ALPL | alkaline phosphatase, liver/bone/kidney | -- | ALPL | -- | Plasma Membrane | phosphatase | 249 | 11647 |
| ANK2 | ankyrin 2, neuronal | ANK2 | -- | -- | Plasma Membrane | other | 287 | 109676 |
| ANKRD11 | ankyrin repeat domain 11 | ANKRD11 | -- | -- | Nucleus | other | 29123 | 77087 |
| ANXA2 | annexin A2 | ANXA2 | -- | -- | Plasma Membrane | other | 302 | 12306 |
| APAF1 | apoptotic peptidase activating factor 1 | APAF1 | -- | -- | Cytoplasm | other | 317 | 11783 |
| APBB2 | amyloid beta (A4) precursor protein-binding, | APBB2 | -- | -- | Cytoplasm | other | 323 | 11787 |
| | family B, member 2 | | | | | | | |
| APLP2 | amyloid beta (A4) precursor-like protein 2 | APLP2 | -- | -- | Cytoplasm | other | 334 | 11804 |
| ARHGAP5 | Rho GTPase activating protein 5 | ARHGAP5 | -- | -- | Cytoplasm | enzyme | 394 | 11855 |
| ARHGAP35 | Rho GTPase activating protein 35 | -- | -- | ARHGAP35 | Nucleus | transcription regulator | 2909 | 232906 |
| ARL13B | ADP-ribosylation factor-like 13B | ARL13B | -- | -- | Extracellular Space | other | 200894 | 68146 |
| ARNT | aryl hydrocarbon receptor nuclear translocator | ARNT | -- | -- | Nucleus | transcription regulator | 405 | 11863 |
| ASPA | aspartoacylase | -- | ASPA | -- | Cytoplasm | enzyme | 443 | 11484 |
| ATF2 | activating transcription factor 2 | ATF2 | -- | ATF2 | Nucleus | transcription regulator | 1386 | 11909 |
| ATG7 | autophagy related 7 | ATG7 | -- | -- | Cytoplasm | enzyme | 10533 | 74244 |
| ATP2C1 | ATPase, Ca++ transporting, type 2C, member 1 | ATP2C1 | -- | -- | Cytoplasm | transporter | 27032 | 235574 |
| ATP7A | ATPase, Cu++ transporting, alpha polypeptide | ATP7A | -- | -- | Plasma Membrane | transporter | 538 | 11977 |
| ATP8A2 | ATPase, aminophospholipid transporter, class I, type 8A, member 2 | ATP8A2 | -- | -- | Plasma Membrane | transporter | 51761 | 50769 |
| ATRX | alpha thalassemia/mental retardation syndrome X-linked | ATRX | -- | -- | Nucleus | transcription regulator | 546 | 22589 |
| ATXN2 | ataxin 2 | ATXN2 | -- | ATXN2 | Nucleus | other | 6311 | 20239 |
| BBS5 | Bardet-Biedl syndrome 5 | BBS5 | -- | -- | Cytoplasm | other | 129880 | 72569 |
| BBS9 | Bardet-Biedl syndrome 9 | BBS9 | -- | -- | Extracellular Space | other | 27241 | 319845 |
| BRAF | B-Raf proto-oncogene, serine/threonine kinase | BRAF | -- | -- | Cytoplasm | enzyme | 673 | 109880 |
| BRAT1 | BRCA1-associated ATM activator 1 | -- | -- | BRAT1 | Nucleus | other | 221927 | 231841 |
| BRCA1 | breast cancer 1, early onset | BRCA1 | -- | -- | Nucleus | transcription regulator | 672 | 12189 |
| BRWD3 | bromodomain and WD repeat domain containing 3 | BRWD3 | -- | -- | Other | other | 254065 | 382236 |
| C9 | complement component 9 | C9 | -- | -- | Extracellular Space | other | 735 | 12279 |
| C2CD3 | C2 calcium-dependent domain containing 3 | C2CD3 | -- | -- | Cytoplasm | other | 26005 | 277939 |
| CACNA2D4 | calcium channel, voltage-dependent, alpha 2/delta subunit 4 | CACNA2D4 | CACNA2D4 | -- | Plasma Membrane | ion channel | 93589 | 466912 |
| CACNB4 | calcium channel, voltage-dependent, beta 4 subunit | CACNB4 | -- | -- | Plasma Membrane | ion channel | 785 | 12298 |
| CAMK1D | calcium/calmodulin-dependent protein kinase ID | CAMK1D | -- | -- | Cytoplasm | kinase | 57118 | 227541 |
| CAMTA1 | calmodulin binding transcription activator 1 | CAMTA1 | -- | -- | Other | other | 23261 | 100072 |
| CASK | calcium/calmodulin-dependent serine protein kinase (MAGUK family) | CASK | -- | -- | Plasma Membrane | kinase | 8573 | 12361 |
| CASP6 | caspase 6, apoptosis-related cysteine peptidase | -- | -- | CASP6 | Cytoplasm | peptidase | 839 | 12368 |
| CASP8 | caspase 8, apoptosis-related cysteine peptidase | CASP8 | -- | CASP8 | Nucleus | peptidase | 841 | 12370 |
| CCM2 | cerebral cavernous malformation 2 | -- | CCM2 | -- | Cytoplasm | other | 83605 | 216527 |
| CDK5RAP2 | CDK5 regulatory subunit associated protein 2 | CDK5RAP2 | -- | -- | Cytoplasm | other | 55755 | 214444 |
| CECR2 | cat eye syndrome chromosome region, candidate 2 | -- | -- | CECR2 | Nucleus | other | 27443 | 330409 |
| CFB | complement factor B | CFB | -- | -- | Extracellular Space | peptidase | 629 | 14962 |
| CFLAR | CASP8 and FADD-like apoptosis regulator | CFLAR | -- | CFLAR | Cytoplasm | other | 8837 | 12633 |
| CLCN3 | chloride channel, voltage-sensitive 3 | CLCN3 | -- | -- | Plasma Membrane | ion channel | 1182 | 12725 |
| CLIC4 | chloride intracellular channel 4 | CLIC4 | CLIC4 | -- | Plasma Membrane | ion channel | 25932 | 29876 |
| CLN6 | ceroid-lipofuscinosis, neuronal 6, late infantile, variant | CLN6 | -- | -- | Cytoplasm | other | 54982 | 76524 |
| CNTN4 | contactin 4 | CNTN4 | -- | -- | Plasma Membrane | enzyme | 152330 | 269784 |
| CNTNAP2 | contactin associated protein-like 2 | CNTNAP2 | -- | -- | Plasma Membrane | other | 26047 | 66797 |
| CPA6 | carboxypeptidase A6 | CPA6 | -- | -- | Extracellular Space | peptidase | 57094 | 329093 |
| CRB1 | crumbs family member 1, photoreceptor morphogenesis associated | CRB1 | -- | -- | Plasma Membrane | other | 23418 | 170788 |
| CREB1 | cAMP responsive element binding protein 1 | -- | -- | CREB1 | Nucleus | transcription regulator | 1385 | 12912 |
| CTSS | cathepsin S | -- | -- | CTSS | Cytoplasm | peptidase | 1520 | 13040 |
| CYP19A1 | cytochrome P450, family 19, subfamily A, polypeptide 1 | CYP19A1 | -- | -- | Cytoplasm | enzyme | 1588 | 13075 |
| CYP7B1 | cytochrome P450, family 7, subfamily B, polypeptide 1 | CYP7B1 | -- | -- | Cytoplasm | enzyme | 9420 | 13123 |
| CYSLTR1 | cysteinyl leukotriene receptor 1 | -- | -- | CYSLTR1 | Plasma Membrane | G-protein coupled receptor | 10800 | 58861 |
| DCLK1 | doublecortin-like kinase 1 | DCLK1 | -- | -- | Other | kinase | 9201 | 13175 |
| DDIT3 | DNA-damage-inducible transcript 3 | -- | -- | DDIT3 | Nucleus | transcription regulator | 1649 | 13198 |
| DIXDC1 | DIX domain containing 1 | -- | -- | DIXDC1 | Cytoplasm | other | 85458 | 330938 |
| DNAJC1 | DnaJ (Hsp40) homolog, subfamily C, member 1 | DNAJC1 | -- | -- | Cytoplasm | other | 64215 | 13418 |
| DST | dystonin | DST | -- | -- | Plasma Membrane | other | 667 | 13518 |
| DYSF | dysferlin | DYSF | -- | -- | Plasma Membrane | other | 8291 | 26903 |
| DYX1C1 | dyslexia susceptibility 1 candidate 1 | DYX1C1 | -- | -- | Nucleus | other | 161582 | 67685 |
| EDEM2 | ER degradation enhancer, mannosidase alpha-like 2 | -- | EDEM2 | -- | Cytoplasm | enzyme | 55741 | 108687 |
| EIF2B3 | eukaryotic translation initiation factor 2B, subunit 3 gamma, 58kDa | EIF2B3 | -- | -- | Cytoplasm | other | 8891 | 108067 |
| EIF2B4 | eukaryotic translation initiation factor 2B, subunit 4 delta, 67kDa | -- | -- | EIF2B4 | Cytoplasm | other | 8890 | 13667 |
| EIF4G3 | eukaryotic translation initiation factor 4 gamma, 3 | EIF4G3 | -- | -- | Cytoplasm | translation regulator | 8672 | 230861 |
| ELMO1 | engulfment and cell motility 1 | ELMO1 | -- | -- | Cytoplasm | other | 9844 | 140580 |
| ENTPD1 | ectonucleoside triphosphate diphosphohydrolase 1 | ENTPD1 | -- | -- | Plasma Membrane | enzyme | 953 | 12495 |
| ERBB3 | erb-b2 receptor tyrosine kinase 3 | -- | ERBB3 | -- | Plasma Membrane | kinase | 2065 | 13867 |
| ERBB2IP | erbb2 interacting protein | ERBB2IP | -- | -- | Plasma Membrane | other | 55914 | 59079 |
| ESR1 | estrogen receptor 1 | ESR1 | -- | -- | Nucleus | ligand-dependent nuclear receptor | 2099 | 13982 |
| F2RL1 | coagulation factor II (thrombin) receptor-like 1 | -- | F2RL1 | -- | Plasma Membrane | G-protein coupled receptor | 2150 | 14063 |
| FAF1 | Fas (TNFRSF6) associated factor 1 | FAF1 | -- | -- | Nucleus | other | 11124 | 14084 |
| FAIM2 | Fas apoptotic inhibitory molecule 2 | -- | -- | FAIM2 | Plasma Membrane | other | 23017 | 72393 |
| FBXL20 | F-box and leucine-rich repeat protein 20 | FBXL20 | -- | -- | Cytoplasm | other | 84961 | 72194 |
| FER | fer (fps/fes related) tyrosine kinase | FER | -- | -- | Cytoplasm | kinase | 2241 | 14158 |
| FGD4 | FYVE, RhoGEF and PH domain containing 4 | FGD4 | -- | -- | Cytoplasm | other | 121512 | 224014 |
| FGF12 | fibroblast growth factor 12 | FGF12 | -- | -- | Extracellular Space | other | 2257 | 14167 |
| FSTL1 | follistatin-like 1 | FSTL1 | -- | -- | Extracellular Space | other | 11167 | 14314 |
| FTO | fat mass and obesity associated | -- | FTO | -- | Nucleus | other | 79068 | 26383 |
| FUCA1 | fucosidase, alpha-L- 1, tissue | FUCA1 | -- | FUCA1 | Cytoplasm | enzyme | 2517 | 71665 |
| FXN | frataxin | -- | -- | FXN | Cytoplasm | kinase | 2395 | 14297 |
| GABBR2 | gamma-aminobutyric acid (GABA) B receptor, 2 | GABBR2 | -- | -- | Plasma Membrane | G-protein coupled receptor | 9568 | 242425 |
| GABRA2 | gamma-aminobutyric acid (GABA) A receptor, alpha 2 | GABRA2 | -- | -- | Plasma Membrane | ion channel | 2555 | 14395 |
| GABRG3 | gamma-aminobutyric acid (GABA) A receptor, gamma 3 | GABRG3 | -- | -- | Plasma Membrane | ion channel | 2567 | 14407 |
| GFRA2 | GDNF family receptor alpha 2 | GFRA2 | -- | -- | Plasma Membrane | transmembrane receptor | 2675 | 14586 |
| GLRA3 | glycine receptor, alpha 3 | GLRA3 | -- | -- | Plasma Membrane | ion channel | 8001 | 110304 |
| GNAQ | guanine nucleotide binding protein (G protein), q polypeptide | GNAQ | -- | -- | Plasma Membrane | enzyme | 2776 | 14682 |
| GOLPH3 | golgi phosphoprotein 3 (coat-protein) | GOLPH3 | -- | -- | Cytoplasm | other | 64083 | 66629 |
| GPHN | gephyrin | GPHN | -- | -- | Plasma Membrane | enzyme | 10243 | 268566 |
| GRID1 | glutamate receptor, ionotropic, delta 1 | GRID1 | -- | -- | Plasma Membrane | G-protein coupled receptor | 2894 | 14803 |
| HDAC2 | histone deacetylase 2 | -- | -- | HDAC2 | Nucleus | transcription regulator | 3066 | 15182 |
| HDAC4 | histone deacetylase 4 | -- | -- | HDAC4 | Nucleus | transcription regulator | 9759 | 208727 |
| HDAC8 | histone deacetylase 8 | HDAC8 | -- | HDAC8 | Nucleus | transcription regulator | 55869 | 70315 |
| HEY2 | hes-related family bHLH transcription factor with YRPW motif 2 | -- | -- | HEY2 | Nucleus | transcription regulator | 23493 | 15214 |
| HLA-DRB1 | major histocompatibility complex, class II, DR beta 1 | HLA-DRB1 | -- | -- | Other | other | 3123 | |
| HLA-E | major histocompatibility complex, class I, E | -- | -- | HLA-E | Plasma Membrane | transmembrane receptor | 3133 | 667803\|15040 |
| HMBOX1 | homeobox containing 1 | -- | HMBOX1 | -- | Nucleus | transcription regulator | 79618 | 219150 |
| HTR1E | 5-hydroxytryptamine (serotonin) receptor 1E, G protein-coupled | HTR1E | -- | -- | Plasma Membrane | G-protein coupled receptor | 3354 | 107927 |
| HTT | huntingtin | HTT | -- | -- | Cytoplasm | transcription regulator | 3064 | 15194 |
| HUWE1 | HECT, UBA and WWE domain containing 1, E3 ubiquitin protein ligase | HUWE1 | -- | -- | Nucleus | transcription regulator | 10075 | 59026 |
| HYOU1 | hypoxia up-regulated 1 | -- | HYOU1 | -- | Cytoplasm | other | 10525 | 12282 |
| IFT140 | intraflagellar transport 140 | IFT140 | -- | -- | Extracellular Space | other | 9742 | 106633 |
| IGBP1 | immunoglobulin (CD79A) binding protein 1 | -- | IGBP1 | -- | Cytoplasm | phosphatase | 3476 | 18518 |
| IL1RAPL1 | interleukin 1 receptor accessory protein-like 1 | IL1RAPL1 | -- | -- | Plasma Membrane | transmembrane receptor | 11141 | 331461 |
| IL1RN | interleukin 1 receptor antagonist | -- | IL1RN | -- | Extracellular Space | cytokine | 3557 | 16181 |
| JARID2 | jumonji, AT rich interactive domain 2 | JARID2 | -- | -- | Nucleus | transcription regulator | 3720 | 16468 |
| KALRN | kalirin, RhoGEF kinase | KALRN | -- | -- | Cytoplasm | kinase | 8997 | 545156 |
| KCNJ6 | potassium channel, inwardly rectifying subfamily J, member 6 | KCNJ6 | -- | -- | Plasma Membrane | ion channel | 3763 | 16522 |
| KDM6A | lysine (K)-specific demethylase 6A | KDM6A | -- | -- | Nucleus | other | 7403 | 22289 |
| KIF15 | kinesin family member 15 | KIF15 | -- | -- | Nucleus | other | 56992 | 209737 |
| LAMA2 | laminin, alpha 2 | LAMA2 | -- | -- | Extracellular Space | other | 3908 | 16773 |
| LEF1 | lymphoid enhancer-binding factor 1 | LEF1 | -- | -- | Nucleus | transcription regulator | 51176 | 16842 |
| LEPR | leptin receptor | LEPR | -- | LEPR | Plasma Membrane | transmembrane receptor | 3953 | 16847 |
| LFNG | LFNG O-fucosylpeptide 3-beta-N-acetylglucosaminyltransferase | -- | -- | LFNG | Cytoplasm | enzyme | 3955 | 16848 |
| LILRB3 | leukocyte immunoglobulin-like receptor, subfamily B (with TM and ITIM domains), 3 | -- | -- | LILRB3 | Plasma Membrane | transmembrane receptor | 11025 | 18733\|18726\|18729\|18725\|18722\|100038909\|100041146 |
| LOC440040 | glutamate receptor, metabotropic 5 pseudogene | LOC440040 | -- | -- | Other | other | 440040 | |
| LPAR4 | lysophosphatidic acid receptor 4 | -- | -- | LPAR4 | Plasma Membrane | other | 2846 | 78134 |
| LPIN1 | lipin 1 | LPIN1 | -- | -- | Nucleus | phosphatase | 23175 | 14245 |
| LRP2 | low density lipoprotein receptor-related protein 2 | LRP2 | -- | -- | Plasma Membrane | transporter | 4036 | 14725 |
| LRP4 | low density lipoprotein receptor-related protein 4 | LRP4 | -- | -- | Extracellular Space | other | 4038 | 228357 |
| LRPAP1 | low density lipoprotein receptor-related protein associated protein 1 | -- | LRPAP1 | -- | Plasma Membrane | transmembrane receptor | 4043 | 16976 |
| LRRC4C | leucine rich repeat containing 4C | LRRC4C | -- | -- | Plasma Membrane | other | 57689 | 241568 |
| LRRK2 | leucine-rich repeat kinase 2 | LRRK2 | -- | -- | Cytoplasm | kinase | 120892 | 66725 |
| MAPK1 | mitogen-activated protein kinase 1 | MAPK1 | -- | -- | Cytoplasm | kinase | 5594 | 26413 |
| MFN2 | mitofusin 2 | MFN2 | -- | -- | Cytoplasm | enzyme | 9927 | 170731 |
| MKKS | McKusick-Kaufman syndrome | -- | -- | MKKS | Cytoplasm | other | 8195 | 59030 |
| MKL1 | megakaryoblastic leukemia (translocation) 1 | MKL1 | -- | -- | Nucleus | transcription regulator | 57591 | 223701 |
| MKS1 | Meckel syndrome, type 1 | MKS1 | -- | -- | Cytoplasm | other | 54903 | 380718 |
| MOS | v-mos Moloney murine sarcoma viral oncogene homolog | -- | -- | MOS | Cytoplasm | kinase | 4342 | 17451 |
| MS4A1 | membrane-spanning 4-domains, subfamily A, | -- | MS4A1 | -- | Plasma Membrane | other | 931 | 12482 |
| | member 1 | | | | | | | |
| MSH2 | mutS homolog 2 | MSH2 | -- | -- | Nucleus | enzyme | 4436 | 17685 |
| MTOR | mechanistic target of rapamycin (serine/threonine kinase) | MTOR | -- | -- | Nucleus | kinase | 2475 | 56717 |
| MYH9 | myosin, heavy chain 9, non-muscle | -- | -- | MYH9 | Cytoplasm | other | 4627 | 17886 |
| MYH10 | myosin, heavy chain 10, non-muscle | MYH10 | -- | -- | Cytoplasm | other | 4628 | 77579 |
| MYO1B | myosin IB | -- | -- | MYO1B | Cytoplasm | other | 4430 | 17912 |
| MYO5A | myosin VA (heavy chain 12, myoxin) | MYO5A | -- | -- | Cytoplasm | enzyme | 4644 | 17918 |
| NBAS | neuroblastoma amplified sequence | NBAS | -- | -- | Nucleus | other | 51594 | 71169 |
| NCOR1 | nuclear receptor corepressor 1 | NCOR1 | -- | -- | Nucleus | transcription regulator | 9611 | 20185 |
| NDUFA9 | NADH dehydrogenase (ubiquinone) 1 alpha subcomplex, 9, 39kDa | -- | -- | NDUFA9 | Cytoplasm | enzyme | 4704 | 66108 |
| NEU1 | sialidase 1 (lysosomal sialidase) | -- | -- | NEU1 | Cytoplasm | enzyme | 4758 | 18010 |
| NFATC3 | nuclear factor of activated T-cells, cytoplasmic, calcineurin-dependent 3 | NFATC3 | -- | -- | Nucleus | transcription regulator | 4775 | 18021 |
| NIPBL | Nipped-B homolog (Drosophila) | NIPBL | -- | -- | Nucleus | transcription regulator | 25836 | 71175 |
| NPEPPS | aminopeptidase puromycin sensitive | NPEPPS | -- | -- | Cytoplasm | peptidase | 9520 | 19155 |
| NPM1 | nucleophosmin (nucleolar phosphoprotein B23, numatrin) | NPM1 | -- | -- | Nucleus | transcription regulator | 4869 | 18148 |
| NR1H2 | nuclear receptor subfamily 1, group H, member 2 | -- | -- | NR1H2 | Nucleus | ligand-dependent nuclear receptor | 7376 | 22260 |
| NRCAM | neuronal cell adhesion molecule | NRCAM | -- | -- | Plasma Membrane | other | 4897 | 319504 |
| NRXN1 | neurexin 1 | NRXN1 | -- | -- | Plasma Membrane | transporter | 9378 | 18189 |
| NRXN2 | neurexin 2 | NRXN2 | -- | -- | Plasma Membrane | transporter | 9379 | 18190 |
| NUSAP1 | nucleolar and spindle associated protein 1 | -- | -- | NUSAP1 | Nucleus | other | 51203 | 108907 |
| OCLN | occludin | OCLN | -- | -- | Plasma Membrane | enzyme | 100506658 | 18260 |
| OMP | olfactory marker protein | -- | OMP | -- | Cytoplasm | other | 4975 | 18378 |
| OPHN1 | oligophrenin 1 | OPHN1 | -- | -- | Cytoplasm | other | 4983 | 94190 |
| ORC6 | origin recognition complex, subunit 6 | ORC6 | -- | -- | Nucleus | other | 23594 | 56452 |
| OTOGL | otogelin-like | OTOGL | -- | -- | Other | other | 283310 | 628870 |
| OXT | oxytocin/neurophysin I prepropeptide | -- | OXT | -- | Extracellular Space | other | 5020 | 18429 |
| PARD3 | par-3 family cell polarity regulator | PARD3 | -- | -- | Plasma Membrane | other | 56288 | 93742 |
| PARG | poly (ADP-ribose) glycohydrolase | PARG | PARG | -- | Cytoplasm | enzyme | 8505 | 26430 |
| PARK2 | parkin RBR E3 ubiquitin protein ligase | PARK2 | -- | -- | Cytoplasm | enzyme | 5071 | 50873 |
| PARK7 | parkinson protein 7 | -- | PARK7 | -- | Nucleus | enzyme | 11315 | 57320 |
| PARP1 | poly (ADP-ribose) polymerase 1 | PARP1 | -- | -- | Nucleus | enzyme | 142 | 11545 |
| PBX3 | pre-B-cell leukemia homeobox 3 | PBX3 | -- | -- | Nucleus | transcription regulator | 5090 | 18516 |
| PCMT1 | protein-L-isoaspartate (D-aspartate) O-methyltransferase | PCMT1 | -- | -- | Cytoplasm | enzyme | 5110 | 18537 |
| PCSK5 | proprotein convertase | PCSK5 | -- | -- | Extracellula | peptidase | 5125 | 18552 |
| | subtilisin/kexin type 5 | | | | r Space | | | |
| PDE11A | phosphodiesterase 11A | PDE11A | -- | -- | Cytoplasm | enzyme | 50940 | 241489 |
| PDE4D | phosphodiesterase 4D, cAMP-specific | PDE4D | -- | -- | Other | enzyme | 5144 | 238871 |
| PDE9A | phosphodiesterase 9A | -- | -- | PDE9A | Cytoplasm | enzyme | 5152 | 18585 |
| PDGFRA | platelet-derived growth factor receptor, alpha polypeptide | PDGFRA | -- | -- | Plasma Membrane | kinase | 5156 | 18595 |
| PDS5B | PDS5 cohesin associated factor B | PDS5B | -- | -- | Nucleus | other | 23047 | 100710 |
| PDXDC2P | pyridoxal-dependent decarboxylase domain containing 2, pseudogene | -- | PDXDC2P | -- | Other | other | 283970 | |
| PEX5 | peroxisomal biogenesis factor 5 | PEX5 | -- | -- | Cytoplasm | transmemb rane receptor | 5830 | 19305 |
| PHB | prohibitin | -- | PHB | -- | Nucleus | transcriptio n regulator | 5245 | 18673 |
| PHC2 | polyhomeotic homolog 2 (Drosophila) | -- | PHC2 | -- | Nucleus | other | 1912 | 54383 |
| PHF8 | PHD finger protein 8 | PHF8 | -- | -- | Nucleus | enzyme | 23133 | 320595 |
| PIGL | phosphatidylinositol glycan anchor biosynthesis, class L | PIGL | -- | -- | Cytoplasm | enzyme | 9487 | 327942 |
| PIGV | phosphatidylinositol glycan anchor biosynthesis, class V | -- | -- | PIGV | Cytoplasm | enzyme | 55650 | 230801 |
| PLCB4 | phospholipase C, beta 4 | PLCB4 | -- | -- | Cytoplasm | enzyme | 5332 | 18798 |
| PNKD | paroxysmal nonkinesigenic dyskinesia | -- | -- | PNKD | Nucleus | other | 25953 | 56695 |
| PNPT1 | polyribonucleotide nucleotidyltransferase 1 | PNPT1 | -- | -- | Cytoplasm | enzyme | 87178 | 71701 |
| POLA1 | polymerase (DNA directed), alpha 1, catalytic subunit | POLA1 | -- | -- | Nucleus | enzyme | 5422 | 18968 |
| POR | P450 (cytochrome) oxidoreductase | POR | -- | -- | Cytoplasm | enzyme | 5447 | 18984 |
| PPARD | peroxisome proliferator-activated receptor delta | PPARD | PPARD | -- | Nucleus | ligand-dependent nuclear receptor | 5467 | 19015 |
| PRKG1 | protein kinase, cGMP-dependent, type I | PRKG1 | -- | -- | Cytoplasm | kinase | 5592 | 19091 |
| PRPH2 | peripherin 2 (retinal degeneration, slow) | -- | -- | PRPH2 | Plasma Membrane | transmemb rane receptor | 5961 | 19133 |
| PSEN1 | presenilin 1 | PSEN1 | -- | PSEN1 | Plasma Membrane | peptidase | 5663 | 19164 |
| PTPN9 | protein tyrosine phosphatase, non-receptor type 9 | PTPN9 | -- | -- | Cytoplasm | phosphatas e | 5780 | 56294 |
| PTPN12 | protein tyrosine phosphatase, non-receptor type 12 | -- | PTPN12 | -- | Cytoplasm | phosphatas e | 5782 | 19248 |
| PTPRD | protein tyrosine phosphatase, receptor type, D | PTPRD | -- | -- | Plasma Membrane | phosphatas e | 5789 | 19266 |
| PUS1 | pseudouridylate synthase 1 | -- | PUS1 | -- | Nucleus | enzyme | 80324 | 56361 |
| RAB7A | RAB7A, member RAS oncogene family | RAB7A | -- | RAB7A | Cytoplasm | enzyme | 7879 | 19349 |
| RAD54L | RAD54-like (S. cerevisiae) | -- | RAD54L | -- | Nucleus | enzyme | 8438 | 19366 |
| RANBP2 | RAN binding protein 2 | RANBP2 | -- | -- | Other | other | 5903 | 19386 |
| RBL1 | retinoblastoma-like 1 | RBL1 | -- | RBL1 | Nucleus | transcriptio n regulator | 5933 | 19650 |
| REEP1 | receptor accessory protein 1 | -- | REEP1 | -- | Cytoplasm | other | 65055 | 52250 |
| RGS11 | regulator of G-protein signaling 11 | -- | -- | RGS11 | Plasma Membrane | enzyme | 8786 | 50782 |
| RLBP1 | retinaldehyde binding protein 1 | -- | -- | RLBP1 | Cytoplasm | transporter | 6017 | 19771 |
| RMND1 | required for meiotic nuclear division 1 homolog (S. cerevisiae) | RMND1 | RMND1 | -- | Cytoplasm | other | 55005 | 66084 |
| RNF135 | ring finger protein 135 | RNF135 | -- | -- | Cytoplasm | enzyme | 84282 | 71956 |
| ROBO2 | roundabout, axon guidance receptor, homolog 2 (Drosophila) | ROBO2 | -- | -- | Plasma Membrane | transmemb rane receptor | 6092 | 268902 |
| RORA | RAR-related orphan receptor A | RORA | -- | -- | Nucleus | ligand-dependent nuclear receptor | 6095 | 19883 |
| RPGRIP1L | RPGRIP1-like | RPGRIP1L | -- | -- | Cytoplasm | other | 23322 | 244585 |
| RTTN | rotatin | RTTN | -- | -- | Other | other | 25914 | 246102 |
| SCN8A | sodium channel, voltage gated, type VIII alpha subunit | SCN8A | -- | -- | Plasma Membrane | ion channel | 6334 | 20273 |
| SEC24D | SEC24 family member D | SEC24D | -- | SEC24 D | Cytoplasm | transporter | 9871 | 69608 |
| SERAC1 | serine active site containing 1 | SERAC1 | -- | -- | Extracellula r Space | other | 84947 | 321007 |
| SHANK2 | SH3 and multiple ankyrin repeat domains 2 | SHANK2 | -- | -- | Plasma Membrane | other | 22941 | 210274 |
| SIL1 | SIL1 nucleotide exchange factor | SIL1 | -- | -- | Cytoplasm | transporter | 64374 | 81500 |
| SLC16A2 | solute carrier family 16, member 2 (thyroid hormone transporter) | -- | SLC16A2 | -- | Plasma Membrane | transporter | 6567 | 20502 |
| SLC17A5 | solute carrier family 17 (acidic sugar transporter), member 5 | SLC17A5 | SLC17A5 | -- | Plasma Membrane | transporter | 26503 | 235504 |
| SLC25A12 | solute carrier family 25 (aspartate/glutamate carrier), member 12 | SLC25A1 2 | SLC25A1 2 | SLC25A 12 | Cytoplasm | transporter | 8604 | 78830 |
| SLC30A10 | solute carrier family 30, member 10 | SLC30A1 0 | -- | -- | Other | transporter | 55532 | 226781 |
| SLMAP | sarcolemma associated protein | -- | SLMAP | -- | Plasma Membrane | other | 7871 | 83997 |
| SMAD2 | SMAD family member 2 | -- | -- | SMAD2 | Nucleus | transcriptio n regulator | 4087 | 17126 |
| SMO | smoothened, frizzled class receptor | -- | -- | SMO | Plasma Membrane | G-protein coupled receptor | 6608 | 319757 |
| SNTB2 | syntrophin, beta 2 (dystrophin-associated protein A1, 59kDa, basic component 2) | SNTB2 | -- | -- | Plasma Membrane | other | 6645 | 20650 |
| SNX2 | sorting nexin 2 | SNX2 | -- | -- | Cytoplasm | transporter | 6643 | 67804 |
| SPRN | shadow of prion protein homolog (zebrafish) | -- | -- | SPRN | Other | other | 503542 | 212518 |
| SPTBN1 | spectrin, beta, non-erythrocytic 1 | SPTBN1 | -- | -- | Plasma Membrane | other | 6711 | 20742 |
| SPTLC1 | serine palmitoyltransferase, long chain base subunit 1 | SPTLC1 | -- | -- | Cytoplasm | enzyme | 10558 | 268656 |
| STAR | steroidogenic acute regulatory protein | -- | -- | STAR | Cytoplasm | transporter | 6770 | 20845 |
| SV2C | synaptic vesicle glycoprotein 2C | SV2C | -- | -- | Plasma Membrane | other | 22987 | 75209 |
| SYNE2 | spectrin repeat containing, nuclear envelope 2 | SYNE2 | -- | -- | Nucleus | other | 23224 | 319565 |
| SYTL4 | synaptotagmin-like 4 | SYTL4 | -- | -- | Cytoplasm | transporter | 94121 | 27359 |
| TACC2 | transforming, acidic coiled-coil containing protein 2 | TACC2 | -- | -- | Nucleus | other | 10579 | 57752 |
| TACR3 | tachykinin receptor 3 | -- | TACR3 | -- | Plasma Membrane | G-protein coupled receptor | 6870 | 21338 |
| TAF1 | TAF1 RNA polymerase II, TATA box binding protein (TBP)-associated factor, 250kDa | TAF1 | TAF1 | TAF1 | Nucleus | transcription regulator | 6872 | 270627 |
| TAOK1 | TAO kinase 1 | TAOK1 | -- | -- | Cytoplasm | kinase | 57551 | 216965 |
| TDRD7 | tudor domain containing 7 | -- | -- | TDRD7 | Cytoplasm | other | 23424 | 100121 |
| TENM4 | teneurin transmembrane protein 4 | TENM4 | -- | TENM4 | Plasma Membrane | other | 26011 | 23966 |
| TFAP2B | transcription factor AP-2 beta (activating enhancer binding protein 2 beta) | -- | TFAP2B | -- | Nucleus | transcription regulator | 7021 | 21419 |
| TIMD4 | T-cell immunoglobulin and mucin domain containing 4 | TIMD4 | -- | -- | Plasma Membrane | other | 91937 | 276891 |
| TMC1 | transmembrane channel-like 1 | TMC1 | -- | TMC1 | Plasma Membrane | ion channel | 117531 | 13409 |
| TRA | T cell receptor alpha locus | TRA | -- | -- | Plasma Membrane | transmembrane receptor | 6955 | 21473 |
| TRIOBP | TRIO and F-actin binding protein | -- | -- | TRIOBP | Nucleus | other | 11078 | 110253 |
| TRPM6 | transient receptor potential cation channel, subfamily M, member 6 | TRPM6 | -- | -- | Plasma Membrane | kinase | 140803 | 225997 |
| TRPV1 | transient receptor potential cation channel, subfamily V, member 1 | -- | -- | TRPV1 | Plasma Membrane | ion channel | 7442 | 193034 |
| TRPV3 | transient receptor potential cation channel, subfamily V, member 3 | -- | TRPV3 | -- | Plasma Membrane | ion channel | 162514 | 246788 |
| UBE4B | ubiquitination factor E4B | -- | UBE4B | -- | Cytoplasm | other | 10277 | 63958 |
| UBR1 | ubiquitin protein ligase E3 component n-recognin 1 | UBR1 | -- | UBR1 | Cytoplasm | enzyme | 197131 | 22222 |
| UBR2 | ubiquitin protein ligase E3 component n-recognin 2 | UBR2 | -- | UBR2 | Nucleus | other | 23304 | 224826 |
| UGT1A3 | UDP glucuronosyltransferase 1 family, polypeptide A3 | -- | UGT1A3 | -- | Cytoplasm | enzyme | 54659 | 22236 |
| ULK1 | unc-51 like autophagy activating kinase 1 | -- | -- | ULK1 | Cytoplasm | kinase | 8408 | 22241 |
| ULK4 | unc-51 like kinase 4 | ULK4 | -- | -- | Other | kinase | 54986 | 209012 |
| UNC13B | unc-13 homolog B (C. elegans) | UNC13B | -- | -- | Cytoplasm | other | 10497 | 22249 |
| USF1 | upstream transcription factor 1 | -- | -- | USF1 | Nucleus | transcription regulator | 7391 | 22278 |
| UTRN | utrophin | UTRN | -- | -- | Plasma Membrane | transmembrane receptor | 7402 | 22288 |
| VAMP7 | vesicle-associated membrane protein 7 | VAMP7 | -- | -- | Cytoplasm | transporter | 6845 | 20955 |
| VAPB | VAMP (vesicle-associated membrane protein)-associated protein B and C | VAPB | -- | -- | Plasma Membrane | other | 9217 | 56491 |
| VAV2 | vav 2 guanine nucleotide exchange factor | VAV2 | -- | -- | Cytoplasm | transcription regulator | 7410 | 22325 |
| VEZT | vezatin, adherens junctions transmembrane protein | VEZT | -- | -- | Plasma Membrane | other | 55591 | 215008 |
| VPS35 | vacuolar protein sorting 35 homolog (S. cerevisiae) | -- | VPS35 | VPS35 | Cytoplasm | transporter | 55737 | 65114 |
| WDR62 | WD repeat domain 62 | WDR62 | -- | WDR62 | Nucleus | other | 284403 | 233064 |
| XK | X-linked Kx blood group | XK | -- | -- | Plasma Membrane | transporter | 7504 | 22439 |
| YARS | tyrosyl-tRNA synthetase | YARS | -- | -- | Cytoplasm | enzyme | 8565 | 107271 |
| YWHAE | tyrosine 3- | -- | YWHAE | -- | Cytoplasm | other | 7531 | 22627 |

| | | | | | | | | | |
|---|---|---|---|---|---|---|---|---|---|
| | monooxygenase/tryptophan 5-monooxygenase activation protein, epsilon | | | | | | | | |
| ZDHHC9 | zinc finger, DHHC-type containing 9 | -- | ZDHHC9 | -- | Cytoplasm | enzyme | 51114 | 208884 |
| ZDHHC15 | zinc finger, DHHC-type containing 15 | ZDHHC15 | -- | -- | Cytoplasm | enzyme | 158866 | 108672 |
| ZEB1 | zinc finger E-box binding homeobox 1 | ZEB1 | -- | -- | Nucleus | transcription regulator | 6935 | 21417 |

**Table S3.** Enrichment of SVA-associated gene functions in IPA knowledge base categories. Functional groups directly attributed to neuronal function are highlighted. **A.** Canonical pathways. **B.** Functional Networks. **C**. Diseases and Functions.

| Ingenuity Canonical Pathways | p-value | A. Molecules in Pathways |
|---|---|---|
| Cardiac β-adrenergic Signaling | 1.04E-01 | CACNA1D,PDE3A,PPP2R3B,CACNA1C,AKAP6,PPP1R3A,PDE4B,PPP1R14B,PDE4D,PPP2R5A,AKAP13,ADCY1, PDE8B,GNG2,PPP2R5E,GNG12 |
| **Synaptic Long Term Potentiation** | 1.14E-01 | MAPK1,GRM1,GRM3,CACNA1C,PPP1R3A,GRM4,CREB5,PPP1R14B,ADCY1,PRKCE,PRKD3,CAMK2B,GRIA3 |
| **Synaptic Long Term Depression** | 9.46E-02 | NOS1,GRID1,MAPK1,GRM3,GRM1,PLA2R1,GUCY2F, PRKCE,PPP2R3B,PPP2R5E,GRM4,PRKD3,PPP2R5A,GRIA3 |
| **Axonal Guidance Signaling** | 6.98E-02 | FYN,LRRC4C,GLI2,MAPK1,EPHB2,NFATC3,PIK3R1, SEMA4F,WNT2,EIF4E,EPHB1,WASL,SRGAP1,UNC5D, PRKCE, ROBO2,SHANK2,PRKD3,GNG12,GLIS1, ITGB1, SEMA3E, SRGAP3,SEMA6D,FZD3,PIK3CB,GNG2, SEMA4B, WNT5A,NRP1 |
| **Neuropathic Pain Signaling In Dorsal Horn Neurons** | 1.07E-01 | CAMK1D,MAPK1,GRM1,GRM3,PIK3R1,PRKCE,KCNQ3, PIK3CB,GRM4,PRKD3,GRIA3,CAMK2B |
| **CREB Signaling in Neurons** | 7.96E-02 | MAPK1,GRM1,GRM3,PIK3R1,GRIK3,GRM4,CREB5, GRID1,ADCY1,PRKCE,PIK3CB,GNG2,PRKD3,GNG12,CAMK2B, GRIA3 |
| **Amyotrophic Lateral Sclerosis Signaling** | 9.17E-02 | CAPN5,NOS1,CACNA1D,GRID1,HECW1,PIK3R1,GRIK3, CACNA1C,PIK3CB,NAIP,GRIA3 |
| TR/RXR Activation | 1.04E-01 | RAB3B,AKR1C1/AKR1C2,SLC2A1,PIK3R1,NCOA1,THRA, PIK3CB,THRB,NCOA3,PPARGC1A |
| Hypoxia Signaling in the Cardiovascular System | 1.23E-01 | BIRC6,UBE2V2,UBE2E2,UBE2E3,CREB5,UBE2E1,PTEN, ARNT |
| Breast Cancer Regulation by Stathmin1 | 7.69E-02 | CAMK1D,MAPK1,PIK3R1,PPP2R3B,PPP1R3A,TRPC5, PPP1R14B,PPP2R5A,ADCY1,PRKCE,PIK3CB,GNG2, PPP2R5E,PRKD3,GNG12,CAMK2B |
| Renal Cell Carcinoma Signaling | 1.08E-01 | MET,ETS1,SLC2A1,MAPK1,CUL2,PIK3R1,PIK3CB,ARNT |
| PKCθ Signaling in T Lymphocytes | 7.75E-02 | MAP3K15,FYN,POU2F1,MAPK1,NFATC3,VAV3,PIK3R1, CD86,PIK3CB,CARD11,CAMK2B |
| **Dopamine-DARPP32 Feedback in cAMP Signaling** | 7.49E-02 | NOS1,CACNA1D,CSNK1G3,CACNA1C,PPP2R3B,PPP1R3A,CREB5, PPP1R14B,PPP2R5A,ADCY1,PRKCE,KCNJ6, PPP2R5E,PRKD3 |
| **Glutamate Receptor Signaling** | 1.01E-01 | GRID1,GRM1,GRM3,GRIK3,GRM4,GNG2,GRIA3 |
| **AMPK Signaling** | 7.23E-02 | PFKFB3,KAT2B,ACACB,MAPK1,PIK3R1,PRKAA2,PPP2R3B,PIK3CB, CFTR,PPP2R5E,AK2,PPP2R5A |
| Actin Cytoskeleton Signaling | 7.05E-02 | ITGB1,MPRIP,MAPK1,PIK3R1,FGF14,TRIO,WASL,PIP5K1C,VAV3, DIAPH2,PIK3CB,WASF2,SSH2,NCKAP1L, ARHGAP24,GNG12,IQGAP3 |
| Clathrin-mediated Endocytosis Signaling | 7.69E-02 | ITGB1,MYO6,AP2B1,APOB,EPS15,EPHB2,PIK3R1,FGF14, ITGB8,HIP1,MET,WASL,PIP5K1C,PIK3CB,MYO1E |
| ERK/MAPK Signaling | 7.35E-02 | PPARG,ITGB1,ETS1,FYN,MAPK1,PIK3R1,PPP2R3B, PPP1R3A,CREB5,PPP1R14B,PPP2R5A,EIF4E,PRKCE, PIK3CB,PPP2R5E |
| Virus Entry via Endocytic Pathways | 9E-02 | ITGB1,FLNB,AP2B1,FYN,PIK3R1,PRKCE,PIK3CB,ITGB8, PRKD3 |
| Fcγ Receptor-mediated Phagocytosis in Macrophages and Monocytes | 8.82E-02 | FYN,MAPK1,VAV3,PIK3R1,DGKB,PRKCE,FYB, PRKD3, PTEN |
| Antiproliferative Role of Somatostatin Receptor 2 | 9.86E-02 | NOS1,MAPK1,PIK3R1,GUCY2F,PIK3CB,GNG2,GNG12 |
| **cAMP-mediated signaling** | 7.31E-02 | MAPK1,CAMK1D,GRM3,PDE3A,AKAP6,PDE4B,GRM4, CREB5,PDE4D,CNGA1,AKAP13,FSHR,ADCY1,P2RY12, PDE8B,CAMK2B |
| Production of Nitric Oxide and Reactive Oxygen Species in Macrophages | 6.67E-02 | MAP3K15,APOB,MAPK1,PIK3R1,PPP2R3B,PPP1R3A, NCF4,PPP1R14B,PPP2R5A,RHOT1,PRKCE,PIK3CB, PPP2R5E,PRKD3 |
| **Glioma Signaling** | 8.04E-02 | CAMK1D,MAPK1,PIK3R1,PRKCE,PIK3CB,RBL1,PRKD3, PTEN,CAMK2B |
| DNA Double-Strand Break Repair by Homologous Recombination | 1.76E-01 | ATRX,RAD52,BRCA1 |
| **Semaphorin Signaling in Neurons** | 1.15E-01 | ITGB1,MET,FYN,MAPK1,RHOT1,NRP1 |
| Macropinocytosis Signaling | 9.21E-02 | ITGB1,MET,PIK3R1,PRKCE,PIK3CB,ITGB8,PRKD3 |
| Relaxin Signaling | 7.01E-02 | MAPK1,PDE3A,PIK3R1,ADCY1,GUCY2F,PDE8B,PIK3CB, PDE4B,GNG2,PDE4D,GNG12 |
| Rac Signaling | 7.38E-02 | ITGB1,NOX4,MAPK1,PIP5K1C,PIK3R1,PIK3CB,PARD3, IQGAP3,ANK1 |
| CDK5 Signaling | 8.51E-02 | ITGB1,MAPK1,ADCY1,PPP2R3B,PPP1R3A,PPP2R5E, PPP1R14B,PPP2R5A |
| CDK5 Signaling | 8.51E-02 | ITGB1,MAPK1,ADCY1,PPP2R3B,PPP1R3A,PPP2R5E, PPP1R14B,PPP2R5A |

| | | | |
|---|---|---|---|
| Cellular Assembly and Organization, Cellular Function and Maintenance, Nucleic Acid Metabolism | AGPAT2,ALMS1,AP1AR,ARHGAP23,ATP5A1,ATP5B, ATPAF1,CGRRF1,CREB3,EDEM2,EXOC6B,EXPH5,FILIP1, GGA1 (includes EG:106039), HSF1,JMJD6,MEPCE, MSTO1,NR1H3,PHRF1,RAB27A,RAD23A,RAE1,RALA, RPAP1,RPS3A,S100PBP,SLC30A9,SMAD9,UBC,VKORC1L1,WDR5,WDR48,WDR59,WDR70 | 17 | 16 |
| Cell Cycle, Hereditary Disorder, **Neurological Disease** | AKAP11,CCHCR1,CNOT8,DEGS1,EIF5,ESF1,FAM21A, IQGAP3,KIAA0196,KIAA1033,KLHDC4,KNTC1,LEPRE1, NBAS,POLD2,PRR12,PUM1,RINT1,SCFD2,SEC61A1, SEC61A2,SEC61G,SERPINA12,SLC39A14,STT3A,TSSC1, TUSC3,TXNDC12,UBC,USP3,WASH1/WASH5P,WDTC1, YTHDF2,ZW10,ZWILCH | 17 | 16 |
| Lipid Metabolism, Nucleic Acid Metabolism, Small Molecule Biochemistry | ACAD9,ACAD10,ACADSB,AGPAT3,AGPAT4,AGPAT5, AGPAT6,CLPP,DDX46,DDX47,EFTUD1,FRYL,GFM2, GPR107,GPR108,GPR119,GPR155,GPR158,GPR176, GPR179,GPR89A/GPR89B,GPR89C,GPRC5C,HECW1, ITFG1,LGR4,NTSR1,OLA1,PCNX,SCAF11,SH2D4A,SUCLG2, TRAPPC12,UBC,ZNF277 | 17 | 16 |
| Organ Morphology, Reproductive System Development and Function, Tissue Morphology | ADAMTS19,ALDH5A1,AQP8,AR,CRISP1,CUX2,DMRT1, DMRTC2,DMRTC1/DMRTC1B,EPB41L4B,FAM110B, FGFR2,GLCCI1,GLIPR1,GPR65,GUCY2F,HOXD1,HS6ST2, KBTBD11,KLK11,KLK14,MED12L,MSMB,MTL5,NLRP10 ,NR3C1,PATZ1,PHF3,RAD54L2,RBMS3,SAR1B,SEMA5B, SPINLW1,TGFB1 (includes EG:21803),TPST2 | 17 | 16 |

| Categories | Diseases or Functions Annotation | p-Value | B. Molecules in Diseases and Functions |
|---|---|---|---|
| Cardiovascular Disease | coronary artery disease | 1.77E-06 | AGPAT5,AKAP13,ASTN2,CACNA1C,CACNA1D, CACNA2D1,CAMTA1,CNTN5,CSMD1,CSMD2, CTNNBL1,EIF4G3,ESR2,GABRA2,GABRB1, KDM1A,KIAA0319L,LRP1 (includes EG:16971), MEF2A,OGDH,P2RY12,PDE3A,PDE4B, PDE4D, PPARG,SGCD,TMEM163,WWOX |
| Endocrine System Development and Function | glucose tolerance | 1.42E-05 | ABCC9,ARNT,CACNA1D,CADPS2,CASP8,CREB5, EIF2AK3,FAM3B,LRP1 (includes EG:16971),MGAT4A,NOS1,PAM,PDK4,PIK3CB, PIK3R1,PPARG,PPARGC1A,PPP1R3A,PRKAA2, PTEN,RPH3AL,SERPINA12,SGK3,SLC30A8, THRA (includes EG:21833) |
| Cell Morphology, Cellular Assembly and Organization, Cellular Development, Cellular Function and Maintenance, Embryonic Development, **Nervous System Development and Function**, Tissue Development, Tissue Morphology | morphology of dendritic spines | 1.93E-05 | CTNNA2,EPHB1,EPHB2,LRRC7,MAGI2,OPHN1 |
| Cellular Movement, **Nervous System Development and Function** | guidance of axons | 6.93E-05 | ALCAM,APBB2,CDH4,CNTN4,EPHB1,EPHB2,FEZ2, GLI2,LRRC4C,MAPK1,NRP1 (includes EG:18186),OPHN1,PLXNA4,RELN,ROBO2,RUNX3, SEMA4F,VAV3 |
| Cellular Function and Maintenance | function of enterocytes | 7.37E-05 | ESR2,MYD88,THRA (includes EG:21833) |
| Connective Tissue Disorders | Dupuytren contracture | 1.69E-04 | ADAMTS16,ADAMTS18,ADAMTS19,ADAMTS3, ADAMTS6,ADAMTS9,COL14A1, COL19A1 (includes EG:12823),COL21A1,COL22A1,COL24A1 |
| **Nervous System Development and Function** | development of central nervous system | 1.70E-04 | ABR,APBB2,ATG7,ATRX,CDCA7L,CTNNA2,DAB1, EPHB2,ESR2,FOXJ3,FYN,FZD3,GLI2,HESX1, IL1RAPL1,IL1RAPL2,KHDRBS1,KIF2A,MAPK1, MCPH1 (includes EG:244329),MET,NDST3, NDUFS4,NPAS2,NPAS3, PARD3,PCSK6,PLXNA4, PPARG,PTEN,RBL1,RELN, REST,ROBO2,RORA, SHROOM4,SLC4A7,SOX6, SPTBN1,ST8SIA4, STK4,THRA (includes EG:21833),TOP2B,TRIO, VSNL1,WASF2,WNT5A, ZEB1 |
| Cardiovascular Disease, Organismal Injury and Abnormalities | variant angina | 1.88E-04 | CACNA1C,CACNA1D,CACNA2D1,SGCD |
| Cell-To-Cell Signaling and Interaction | synaptic depression | 2.24E-04 | ADCY1,CAMK2B,EPHB1,EPHB2,GRIA3,GRM1, GRM3,HIP1 (includes EG:176224), LRRC7,PCLO, PIP5K1C, ST8SIA4,SYN3 |
| **Neurological Disease** | radiculopathy | 2.85E-04 | CACNA2D1,CACNA2D2,CACNA2D3 |
| Organismal Survival | organismal death | 3.14E-04 | ABCC9,ACLY,ADAR,ANKH,ANXA1,APBB2,APOB, ARNT,ATG7,BIRC6,BMPR1A,BRCA1,BRD4, C4B (includes others),CACNA2D2,CASP8, CELF1,CFTR,CLIC4, CUZD1,DAB1,DMD,EHMT2 ,EIF2AK3,ENG,EPHB2, ESR2,ETS1,ETV6,FBN1,FLNB, FYN,GABRA2,GPR98, HIP1 (includes EG:176224),HMGA2, IL4R,IRGM,ITGB1,ITGB8, KAT2B,KCNJ6,KIDINS220, LRRC7,MAD1L1,MADD, MAGI2,MAPK1,MCPH1 (includes EG:244329), MEF2A,MET,MGMT, MYBL2,MYD88,NCOA1, NCOA3,NDUFS4,NFATC3,NOS1,NOVA1,NRP1 (includes EG:18186),OPHN1, PARP1,PCLO, PCSK6,PHC1,PHC2,PIK3CB,PIK3R1, PIP5K1C, POU2F1,PPARG,PPARGC1A,PPARGC1B, PTEN, PTPRJ,RAB3B,RAD52 (includes EG:19365), RBL1, REST, REV3L, SGCD,SIAH1,SLC2A1, SPEN, SPRED2, SPTBN1,ST8SIA4,STK4,TCF12,TFCP2L1,TFEB, THRA (includes EG:21833),THRB,TOP2B, TPMT, TRAFD1, TRIM54,TRIO,UBP1 (includes EG:100136855), USP18,VAV3,WASF2,WASL,WWOX,ZEB1 |
| **Cellular Function and Maintenance, Nervous System** Development and Function | function of outer hair cells | 3.27E-04 | CACNA1D,GPR98,THRA (includes EG:21833),THRB |
| Cancer, Hereditary Disorder, Reproductive System Disease | tumorigenesis of prostatic carcinoma | 3.27E-04 | HIP1 (includes EG:176224),PPARG,PRKCE,PTEN |
| DNA Replication, Recombination, and Repair | breakage of DNA | 3.66E-04 | BRCA1,NOS1,PARP1,PTEN,RAD52 (includes EG:19365),REV3L,TOP2B |
| Cardiovascular Disease, Organismal Injury and Abnormalities | angina pectoris | 4.31E-04 | CACNA1C,CACNA1D,CACNA2D1,P2RY12,PDE3A, PDE4B,PDE4D,PPARG,SGCD |
| Cellular Assembly and Organization, Cellular Function and Maintenance | organization of cytoplasm | 4.49E-04 | ABR,ADCY1,ALMS1,ANTXR1,ARHGAP24, ARHGAP32,ARHGAP6,ARHGAP8/PRR5-ARHGAP8,ATG7,ATP7A,CAMK1D,CDH4,CNTN4, CTNNA2,CTNND2,DAB1,DIAPH2,EIF2AK3,ENG, EPHB1,EPHB2,FHIT,FLNB,FYN,GRM4,HYDIN,ITGB1,ITGB8,KIDINS220,KIF20B,KIF2A,LRP1 (includes EG:16971),LRRC4C,LRRC7, MAGI2, MAP4,MAPK1, MET,MPRIP,MSTO1,MYD88, MYO6,NCKAP1L, NOS1,NPLOC4,NRP1 (includes EG:18186), OPHN1,PACSIN2,PAM,PARD3, PCDH15, PCLO, PIK3R1,PIP5K1C,PKHD1, PLXNA4,PPARG, PPARGC1A, PPARGC1B, PRKCE,PTEN,PTPRQ,RALB, RELN, ROBO2, RRBP1,RUNX3,SEC16B,SEMA3E, SEMA4F, SEPT7,SEPT9,SHROOM3, SHROOM4,SIAH1, SSH2,STRN,TFCP2L1,TOP2B, TRIM54,TRIO,TTK, VAV3, VPS54, WASF2,WASL,WNT5A |
| **Nervous System Development and Function** | morphology of nervous system | 5.03E-04 | ABR,ADCY1,ALPL,APBB2,APOB,BRCA1, CACNA1D, CACNA2D2,CADPS2,CD86,CLIC4, CNTN4,CTNNA2 ,CUX2,DAB1,ECE1,EPHB1, EPHB2, ERBB2IP,ESR2, FAT3,FSHR,FYN,GLI2, GRID1, HESX1,IL1RAPL1,IL4R, ITGB1,ITGB8, KIDINS220,KIF2A,LRRC7,MADD, MAGI2,MCPH1 (includes EG:244329),MET,NCOA1, NDST3, NDUFS4,NFATC3,NOS1,NPAS3,NRP1 (includes EG:18186),OPHN1 ,PCSK6,PLXNA4, PPARGC1A, PTEN,RBL1,RELN,ROBO2,SEMA5B,ST8SIA4,THRA (includes EG:21833), THRB,TOP2B,TRIO, VAV3, WNT5A |
| Molecular Transport | transport of Mg2+ | 5.28E-04 | MAGT1,NIPAL1,TUSC3,ZDHHC17 |
| **Neurological Disease** | hydrocephalus | 5.75E-04 | APOB,DAB1,ENG,GLI2,ITGB8,OPHN1,PTEN, RELN, ST8SIA4,USP18 |
| Cellular Assembly and Organization, Cellular Function and Maintenance | organization of cytoskeleton | 5.85E-04 | ABR,ADCY1,ALMS1,ANTXR1,ARHGAP24, ARHGAP32,ARHGAP6,ARHGAP8/PRR5-ARHGAP8, ATG7,ATP7A,CAMK1D,CDH4,CNTN4, CTNNA2, CTNND2,DAB1,DIAPH2,ENG,EPHB1, EPHB2, FHIT, FLNB,FYN,GRM4,HYDIN, ITGB1,ITGB8, KIDINS220, KIF20B,KIF2A,LRP1 (includes EG:16971), LRRC4C, LRRC7,MAGI2, MAP4,MAPK1, MET,MPRIP, MYD88, MYO6, NCKAP1L,NOS1, NRP1 (includes EG:18186), OPHN1,PACSIN2, PAM, PARD3,PCDH15,PCLO, PIK3R1,PIP5K1C,PKHD1,PLXNA4,PPARG,PPARGC 1B,PRKCE,PTEN,PTPRQ, RALB,RELN, ROBO2, RRBP1, RUNX3,SEMA3E, SEMA4F,SEPT7,SEPT9, SHROOM3, SHROOM4, SSH2,STRN,TOP2B, TRIM54, TRIO, VAV3,VPS54, WASF2,WASL,WNT5A |
| Cell Cycle, Cellular Movement | cytokinesis of tumour cell lines | 5.86E-04 | BRD4,KIF20B,MASTL,NEDD4L,SEPT7,SEPT9, SIAH1, VAV3 |
| Cardiovascular System Development and Function | abnormal morphology of vasculature | 6.59E-04 | ARNT,BIRC6,CLIC4,FBN1,ITGB8,MAPK1,NDUFS4, NFATC3,NPAS3,PPARG,PTEN,SLC4A7,STK4, UBP1 (includes EG:100136855),WASF2,WNT2 |
| Cardiovascular System Development and Function, Embryonic Development, Organismal Development, Tissue Development | vascularization of yolk sac | 8.02E-04 | ARNT,ITGB8,NRP1 (includes EG:18186),STK4 |
| Cell Cycle, DNA Replication, Recombination, and Repair | abnormal morphology of chromosomes | 9.14E-04 | BRCA1,HELLS,MCPH1 (includes EG:244329), PARP1,RAD52 (includes EG:19365) |
| Infectious Disease | infection by Retroviridae | 1.07E-03 | ACACB,AKAP13,ANKRD30A,ARHGAP32,ATG7, BRCA1,CACNA2D1,CACNA2D2,CACNA2D3, CALCOCO1,CAMK1D,CD86,CFTR,CLOCK,DAPK2, DMXL1,DYSF,EIF2C3,FDPS,GABRB1,HCP5,IL4R, KHDRBS1,KIAA0922,MAP4,NDUFAF2,PCSK6, PDE3A ,PDE4B,PDE4D,PHF12,PHF3,PIP5K1C, PPARG, PPP2R5E,PTPRJ,PVT1,RALB,RANBP17 ,RNF170, SEMA5B,SLC2A1,SLC2A13, SLC4A7, SPEN, SPTBN1, ST3GAL3,SUCLG2, TATDN1, TMEM163, TOP2B, TRIM5,UBE2E1, USP26, WDTC1,XKR4,ZNF292 |
| Developmental Disorder, **Neurological Disease** | mental retardation | 1.15E-03 | ATP7A,ATRX,AUTS2,CTNND2,GRIA3,IL1RAPL1,MAGT1,OPHN1,SHROOM4,TSPAN7,TUSC3,ZNF599 |
| **Behaviour** | nest building behaviour | 1.16E-03 | LRRC7,MAPK1,NDUFS4,NPAS3 |
| Cardiovascular System Development and Function, Organ Morphology | abnormal morphology of heart | 1.23E-03 | ACACB,CASP8,ECE1,ESR2,ITGB8,MAPK1, NFATC3, NOX4,PHC1,PPARGC1A,PPARGC1B, SGCD,SNX27, SPTBN1,TRIM54,UBP1 (includes EG:100136855),WASF2,WASL |
| Digestive System Development and Functio Endocrine System Development and Functio Organ Morphology | morphology of pancreas | 1.25E-03 | CACNA1D,CADPS2,CFTR,EIF2AK3,ESR2,GPR39, MGMT,MYD88,NFATC3,PKHD1,PPARG,PTEN, THRA (includes EG:21833) |

Table S4. Link of SVA-associated genes to neural system related ontological categories. Shading represents an association between genes with particular types of SVAs (columns) with certain categories (rows). An enrichment in a category with (-log(p-value)>2.9) is highlighted in purple and (-log(p-value)> 4.6), in blue.

| SVA type | A | B | C | D | E | F | F1 |
|---|---|---|---|---|---|---|---|
| **Behavioral features** | | | | | | | |
| behavior | - | 3.754903399 88501 | 1.667197019 29576 | 7.697397238 39598 | - | 2.344704794 37986 | 2.23 |
| head tossing | - | - | 2.226348280 65468 | - | - | - | - |
| abnormal initiation of locomotion | - | - | - | 2.186947690 0683 | - | - | - |
| aggressive behavior | - | 3.18572896506994 | - | 2.99108461818918 | - | - | - |
| aggressive behavior toward females | - | - | - | 3.29807623617792 | - | - | - |
| associative memory | - | - | - | - | - | 2.32080527409475 | - |
| auditory brainstem response | - | - | 1.75173307322935 | - | - | - | - |
| barbering behavior | - | 2.08605893635069 | - | - | - | - | - |
| climbing ability | 2.91189286330032 | - | - | - | - | - | - |
| climbing behavior | - | 2.26156736922413 | - | - | - | - | - |
| coordination | - | 2.48681729898497 | - | 3.67182351782147 | - | - | - |
| coordination of limb | - | - | 1.38875560535539 | - | - | - | - |
| discriminatory learning | - | 3.58236301911825 | - | - | 1.98076455827166 | - | - |
| facial expression | - | 2.56201430010179 | - | - | - | - | - |
| feeding | - | - | - | 2.62227467361367 | - | - | 1.88 |
| grooming | - | 3.31565926146307 | 2.25980197067178 | 3.1582288478537 | - | - | - |
| learning | - | 4.7671531581683 | - | 4.97082157222324 | - | - | - |
| limb clasping | - | - | - | - | - | - | 2.2 |
| locomotor activity | - | - | - | - | - | - | 1.48 |
| long-term spatial reference memory | - | - | - | - | - | 2.62132830131439 | - |
| mating | - | - | - | 2.29728747764665 | - | - | - |
| mating behavior | - | - | - | 2.47190514059428 | - | - | - |
| memory | - | 2.45253338698332 | - | 2.20867174581991 | 1.54035438486205 | - | - |
| mother preference | - | - | - | - | - | 2.62132830131439 | - |
| motor learning | - | 4.17316098655729 | - | - | - | - | - |
| mounting | - | - | 2.10182751405373 | - | - | - | - |
| movement of rodents | - | - | - | 2.33021962354048 | 1.5873195307674 | - | - |
| nest building behavior | - | 3.12203289994915 | - | 3.81258793544142 | - | - | - |
| nursing | - | - | - | - | - | - | 1.82 |
| olfactory discrimination | - | 3.48621278845036 | - | - | - | - | - |
| parental behavior | - | - | - | 2.85422866327821 | - | - | - |
| passive avoidance learning | 1.33043003745532 | 2.29158203899342 | - | - | 1.35114076601387 | - | - |
| place preference | - | - | - | - | 1.36218234521848 | - | - |
| pup retrieval | - | - | - | 2.56996747931816 | - | - | - |
| sexual behavior | - | - | - | 3.21841538890373 | - | - | - |
| sexual receptivity of female organism | - | - | - | 3.12403012397743 | - | - | - |
| short-term memory | - | - | - | - | 1.59570755244453 | - | - |
| social behavior | - | 4.47518090165667 | - | 3.9260225846134 | - | 3.26212780530914 | - |
| social exploration | - | 2.31368906798083 | - | - | - | - | - |
| social odor recognition memory | - | - | - | - | - | 2.62132830131439 | - |
| somatosensory-discrimination learning | - | 2.56201430010179 | - | - | - | - | - |
| spatial learning | - | 3.81012718176609 | - | - | 2.89626985782354 | - | - |
| stress response of mice | - | 3.69074757210718 | - | - | - | - | - |
| stretching behavior | - | 2.56201430010179 | - | - | - | - | - |
| vocalization | - | - | - | 3.08064267948855 | - | - | - |
| walking | - | - | - | 3.48658502379951 | - | - | 1.5 |
| **Diseases SVA type** | **A** | **B** | **C** | **D** | **E** | **F** | **F1** |
| Alzheimer's disease | | | 2.9551576643813 | | | 2.52185071754049 | |
| Bipolar disorder | | | | 2.38113792337015 | | | |
| Central nervous system tumor | | | | 4.20745620272873 | | | |
| Cerebellar ataxia | | | | 3.83025436749913 | | | |
| Dementia | | | | 2.74171031910954 | | | |
| Depressive disorder | | | | 3.68983865524415 | | | |
| Epilepsy | | | | 2.67999843154644 | | 2.62132830131439 | |
| Familial Parkinson disease | | | | 2.3574004396226 | | 3.30221792079955 | |
| Generalized seizures | | | | 3.51212927259493 | | | |
| Glioma | | | | 4.36604317311433 | | | |
| Hereditary CNS demyelinating disease | | | | 4.47653541539325 | | | |
| Huntington's Disease | | 2.08605893635069 | | | | | 1.89 |
| Localization-related epilepsy | | | | 2.96498772553459 | | | |
| Malformation of brain | 1.34182843509574 | 7.23986889340811 | 1.57587633855576 | 6.08733393596778 | 2.39801008187586 | 5.39579657247572 | |
| Mental retardation | | 5.01977760752792 | 2.26395923402463 | 8.16037961988198 | | | |
| Mood Disorders | | | | 3.91783633940726 | 2.51841337092726 | | |
| Movement Disorders | | | 8.00533347570778 | 7.87559858904657 | | | 1.96 |
| neuromuscular disease | | | 3.75767432494131 | 3.66763736137556 | 1.48569674713159 | | 2.3 |
| Parkinson disease | | | 2.70263324098405 | 2.14420704661376 | | 2.62132830131439 | |
| Progressive motor neuropathy | | | | 3.74959426346343 | | | |
| Schizophrenia | | 2.56201430010179 | | 2.28225465194161 | | | |
| Seizures | 1.73910092696039 | | | 4.54232933858452 | | | 1.39 |
| syndromic X-linked mental retardation | | 2.56201430010179 | | 5.99085066237035 | 3.38382739638322 | | |
| X linked mental retardation syndromic 60 | | 2.91171969671727 | | 7.10993984472082 | 2.93308216239815 | | |
| anxiety-like behavior | | 3.92502945950205 | | | | | |

| **Biological functions SVA type** | **A** | **B** | **C** | **D** | **E** | **F** | **F1** |
|---|---|---|---|---|---|---|---|
| assembly of axon initial segments | - | - | - | 3.81686130909647 | - | - | - |
| axonogenesis | - | - | - | - | - | 4.18816776502243 | - |
| concentration of norepinephrine | - | - | - | - | - | 3.38200803578627 | - |
| dendritic growth/branching | - | - | - | 4.71750799432237 | - | - | - |
| development of brain | - | 7.98334449039365 | - | 13.4999329586417 | - | 6.54409076147706 | 1.31 |
| development of central nervous system | - | - | - | 16.2694644544585 | - | 9.61478502402455 | - |
| development of forebrain | - | 6.53406412330348 | - | 5.83283644943782 | 1.48964039229991 | 7.01943136500185 | - |
| electrophysiology of nervous system | - | - | - | 4.3396850486171 | - | - | - |
| formation of cellular protrusions | - | 3.47509841194994 | - | 12.2469453253689 | - | 2.5166238536187 | - |
| long-term potentiation | - | 2.37360481813443 | - | 2.75089557668827 | 1.49950612706148 | - | - |
| morphology of axons | - | 4.01024659091482 | - | - | - | - | - |
| morphology of brain | 1.5525181561587 | 5.74463125089947 | 4.08437055195381 | 10.3406916560822 | - | 7.37851670496971 | - |
| morphology of forebrain | - | 5.35318521268852 | - | 4.95124388467946 | - | 4.61540182788052 | - |
| morphology of nervous system | - | 5.9137846070663 | 4.94059567478393 | 17.8900878812047 | 1.6310582301263 | 8.59405181098923 | - |
| morphology of telencephalon | 1.71304474780371 | 5.08935490836635 | 3.40742402197678 | 3.55489802902982 | - | 4.22300864096256 | - |
| neuronal cell death | 2.06491332796842 | - | 2.6997902426981 | 2.91603242078867 | 1.3664435923436 | 2.27753591642938 | - |
| quantity of axons | - | - | - | 5.25752610669274 | - | - | - |
| quantity of brain cells | - | - | - | - | - | 5.04155616270202 | - |
| synaptic depression | - | 2.45817742554262 | - | 5.87621865807706 | - | - | - |
| synaptic transmission | 4.19583462577855 | - | - | - | - | - | - |
| synaptogenesis | - | 5.50608308707257 | - | 4.00174359428815 | - | - | - |
| neurotransmission | - | 4.56702557465991 | - | 2.82319293681402 | - | 2.37748244844698 | - |

**Table S5**. Genes grouped by their associated behavioural categories and the SVA-types (a (SVA_A)-g(SVA_F1)).

**a.SVA_A**

| Diseases or Functions Annotation | Molecules |
|---|---|
| Climbing Ability | KCNJ6 |
| Anxiety | GABBR2,KCNJ6,PLCB4 |
| Passive Avoidance Learning | GABBR2 |

**b.SVA_B**

| Diseases or Functions Annotation | Molecules |
|---|---|
| Learning | HDAC4,HTT,MAPK1,NRCAM,NRXN1,OXT,TACR3,UBR1 |
| Social Behaviour | HTT,MAPK1,NRXN1,OXT |
| Motor Learning | HTT,NRCAM,NRXN1 |
| Anxiety-Like Behaviour | HTT,MAPK1,OXT |
| Spatial Learning | HDAC4,HTT,NRCAM,TACR3,UBR1 |
| Behaviour | CLCN3,HDAC4,HTT,MAPK1,NRCAM,NRXN1,OXT,SEC24D,TACR3,UBR1,F1 |
| Discriminatory Learning | HTT,UBR1 |
| Olfactory Discrimination | HTT,OXT |
| Grooming | HTT,NRXN1,OXT |
| Emotional Behaviour | HTT,MAPK1,NRXN1,OXT,TACR3 |
| Aggressive Behaviour | HTT,MAPK1,OXT |
| Nest Building Behaviour | MAPK1,NRXN1 |
| Invasive Behaviour | NRCAM,SEC24D |
| Vertical Rearing | HTT,NRCAM,UBR1 |
| Circadian Rhythm | HTT,MAPK1,OXT |
| Apathy | HTT |
| Choreiform Movement | HTT |
| Facial Expression | HTT |
| Irritability Of Organism | HTT |
| Somatosensory-Discrimination Learning | HTT |
| Spasmodic Movement | HTT |
| Stretching Behaviour | HTT |
| Memory | HDAC4,HTT,MAPK1,OXT |
| Hyperactivity | HTT,TIMD4 |
| Social Exploration | NRCAM,OXT |
| Passive Avoidance Learning | TACR3,UBR1 |
| Climbing Behaviour | HTT |
| Executive Functioning | HTT |
| Barbering Behaviour | USF1 |

**c.SVA_C**

| **Diseases or Functions Annotation** | **Molecules** |
|---|---|
| Grooming | LRRK2,SLC16A2 |
| Head Tossing | ATF2 |
| Mounting | NPEPPS |
| Emotional Behaviour | LRRK2,MKKS,SLC16A2 |
| Behaviour | ABHD12,ATF2,LRRK2,MKKS,NPEPPS,SLC16A2 |
| Locomotion | ABCA1,FGF12,LRRK2 |

**d.SVA_D**

| **Diseases Or Functions Annotation** | **Molecules** |
|---|---|
| Behaviour | ABL2,ATP8A2,CACNB4,CNTNAP2,CYP19A1,ESR1,F2RL1,FTO, GFRA2,GNAQ,GPHN,IL1RN,LEPR,LRPAP1,MOS,MTOR,NFATC3, NRXN1,OCLN,PARK7,PRKG1,SCN8A,SHANK2,SYTL4, TRPV1,UBR1,VEZT |
| Cognition | ABL2,CNTNAP2,CYP19A1,ESR1,IL1RN,LEPR,LRPAP1,MTOR, NIPBL,NRXN1, PARK7,SHANK2 ,UBR1,VEZT |
| Learning | ABL2,CNTNAP2,CYP19A1,ESR1,IL1RN,LEPR,LRPAP1,MTOR,NRXN1,PARK7, SHANK2,UBR1, VEZT |
| Locomotion | ABL2,AGTPBP1,ATP7A,CACNB4,CYP19A1,ESR1,FXN,NFATC3, PARK7,PBX3, SCN8A |
| Social Behaviour | CNTNAP2,IL1RN,NFATC3,NRXN1,SHANK2 |
| Nest Building Behaviour | CNTNAP2,NRXN1,SHANK2 |
| Walking | AGTPBP1,CACNB4,FXN,SCN8A |
| Emotional Behaviour | ABL2,CNTNAP2,CYP19A1,ESR1,LEPR,NRXN1 ,SHANK2,TRPV1 |
| Aggressive Behaviour Toward Females | CYP19A1,ESR1 |
| Sexual Behaviour | ABL2,CYP19A1,ESR1,SYTL4 |
| Grooming | CNTNAP2,CYP19A1,NRXN1,SHANK2 |
| Sexual Receptivity Of Female | CYP19A1,ESR1 |
| Organism | |
| Vocalization | CNTNAP2,GPHN,SHANK2 |
| Aggressive Behaviour | ABL2,CYP19A1,ESR1,LEPR |
| Parental Behaviour | CYP19A1,ESR1,SHANK2 |
| Feeding | ATP8A2,ESR1,FTO,GPHN,IL1RN,LEPR,OCLN |
| Pup Retrieval | ESR1,SHANK2 |
| Mating Behaviour | ABL2,CYP19A1,ESR1 |
| Anxiety | GABRA2,GABRG3,NFATC3,SHANK2,TRPV1, VAMP7 |
| Mating | CYP19A1,ESR1 |
| Panic-Like Anxiety | GABRA2,GABRG3 |
| Memory | ABL2,CYP19A1,ESR1,IL1RN,LEPR,MTOR |
| Abnormal Initiation Of Locomotion | CYP19A1,NFATC3,PARK7 |

**e. SVA_E**

| Diseases or Functions Annotation | Molecules |
|---|---|
| Spatial Learning | CREB1,OPHN1,UBR1 |
| Vertical Rearing | CREB1,UBR1 |
| Discriminatory Learning | UBR1 |
| Aggressive Behaviour Toward Mice | OPHN1 |
| Anxiety | CREB1,OPHN1 |
| Short-Term Memory | CREB1 |
| Addiction Behaviour | CREB1 |
| Memory | CREB1,OPHN1 |
| Place Preference | CREB1 |
| Passive Avoidance Learning | UBR1 |

**f. SVA_F**

| Diseases or Functions Annotation | Molecules |
|---|---|
| Social Behaviour | KALRN,OMP,PDE11A |
| Long-Term Spatial Reference Memory | PSEN1 |
| Mother Preference | OMP |
| Social Odour Recognition Memory | PDE11A |
| Behaviour | APLP2,KALRN,LEF1,MYH10,OMP,PARK2,PDE11A,PSEN1 |
| Associative Memory | PSEN1 |

**g.SVA_F1**

| Diseases or Functions Annotation | Molecules |
|---|---|
| Behaviour | ACOT11,ATG7,CYP7B1,MKL1 |
| Limb Clasping | ATG7 |
| Feeding | ACOT11,MKL1 |
| Nursing | MKL1 |
| Walking | ATG7 |